\documentclass[english, twocolumn, superscriptaddress, aps, showpacs, showkeys, amsmath, amssymb, floatfix, nofootinbib]{revtex4-1}

\pdfoutput=1
\usepackage[pdftex]{graphicx}
\usepackage[pdftex]{color}
\usepackage[colorlinks,bookmarks]{hyperref}
\definecolor{linkblue}{rgb}{0,0,0.8}
\definecolor{linkgreen}{rgb}{0,0.5,0}
\hypersetup{linkcolor=linkblue, citecolor=linkgreen, urlcolor=linkblue}

\usepackage[T1]{fontenc}
\usepackage[utf8]{inputenc}
\usepackage{lmodern}
\usepackage{babel}
\usepackage{bm}
\usepackage{esint}
\usepackage{verbatim}
\usepackage[us]{datetime}
\usepackage{booktabs}
\usepackage{enumerate}
\usepackage{graphics}
\usepackage{float}
\usepackage{amsmath}
\usepackage{amsfonts}
\usepackage{amssymb}

\usepackage{multirow}
\usepackage{array}
\usepackage[normalem]{ulem}

\def\rd{{\rm d}}
\def\E{\mathcal{E}}
\def\L{\mathcal{L}}
\def\Om{\Omega_{\rm m}}

\def\dc{d_L^{(\rm cut)}}

\def\dccs{d_C^{(\rm cut2)}}
\def\dcct{d_C^{(\rm cut3)}}
\def\Mz{\mathcal{M}_c}

\newcommand{\pa}[1]{\left(#1\right)}
\newcommand{\paq}[1]{\left[#1\right]}

\newcommand{\be}{\begin{equation}}
\newcommand{\ee}{\end{equation}}
\newcommand{\bdm}{\begin{displaymath}}
\newcommand{\edm}{\end{displaymath}}
\newcommand{\ba}{\begin{array}}
\newcommand{\ea}{\end{array}}
\newcommand{\ds}{\displaystyle}

\begin{document}

\title{Measuring the Hubble constant with black sirens}

\author{Hebertt Leandro}
\affiliation{Departamento de F\'\i sica Te\'orica e Experimental,
Universidade Federal do Rio Grande do Norte, 59078-970, Natal-RN, Brazil}
\email{heberttls22@gmail.com}

\author{Valerio Marra}
\affiliation{N\'ucleo de Astrofísica e Cosmologia \& Departamento de F\'\i sica, Universidade Federal do Espírito Santo, 29075-910, Vitória, ES, Brazil,\\
  INAF, Osservatorio Astronomico di Trieste, via Tiepolo 11, 34131, Trieste, Italy,
  IFPU, Institute for Fundamental Physics of the Universe, via Beirut 2, 34151, Trieste, Italy}
\email{valerio.marra@me.com}

\author{Riccardo Sturani}
\affiliation{International Institute of Physics, Universidade Federal do Rio Grande
  do Norte, 59078-970, Natal-RN, Brazil}
\email{riccardo.sturani@ufrn.br}

\begin{abstract}
We investigate a recently proposed method for measuring the Hubble constant from
gravitational wave detections of binary black hole coalescences  without electromagnetic
counterparts. In the absence of a direct redshift measurement, the missing
information on the left-hand side of the Hubble-Lemaître law is
provided by the statistical knowledge on the redshift distribution of
sources. We assume that source distribution in redshift depends on 
unknown hyperparameters, modeling our ignorance of the astrophysical binary
black hole distribution.
With tens of thousands of these ``black sirens'' -- a realistic figure for the third generation detectors Einstein Telescope  and Cosmic Explorer  -- an observational
constraint on the value of the Hubble parameter at percent level can be
obtained. This method has the advantage of not relying on electromagnetic counterparts,
which accompany a very small fraction of gravitational wave detections, nor on often unavailable or incomplete galaxy catalogs.
\end{abstract}


\maketitle

\section{Introduction}\label{intro}

The Hubble constant $H_0$ -- the current expansion rate of space -- is a fundamental parameter that sets the time and distance scales of the observable
Universe.
It is then alarming that the local model-independent determination of the Hubble  constant  via calibrated local Type Ia supernovae
~\cite{Riess:2020fzl}  is in strong tension with the CMB determination based on the standard $\Lambda$CDM model of cosmology
~\cite{Aghanim:2018eyx}.
The tension reached $4.5\sigma$~\cite{Camarena:2021jlr} and it could very well signal the need of a new standard model of cosmology~\cite{Knox:2019rjx}.
The possibility of physics beyond $\Lambda$CDM has been urging the scientific community to measure $H_0$ via the widest range possible of probes and techniques: besides Cepheids, strong lensing time delays, tip of the red giant branch, megamasers, oxygen-rich Miras and surface brightness fluctuations~(see \cite{Verde:2019ivm,DiValentino:2021izs,Perivolaropoulos:2021jda,Khetan:2020hmh} for details). 

Gravitational wave (GW) observations are expected to play an important role in the determination of $H_0$ already in the near future \cite{Gray:2019ksv},
thanks first to the second generation detectors LIGO \cite{TheLIGOScientific:2014jea},
Virgo \cite{TheVirgo:2014hva} and KAGRA \cite{KAGRA:2020tym},
and then to the third generation detectors Einstein Telescope \cite{Punturo:2010zz} and Cosmic Explorer \cite{Evans:2016mbw}.
The reason is twofold. First, GW observations are a new and powerful probe so that an independent and precise measurement of $H_0$ will be obtained.
Second, GW observations already with second generation detectors will cover  the most interesting redshift range ($0.2\lesssim z \lesssim 0.7$,  \citealt{KAGRA:2020npa}) as far as the Hubble tension is concerned.
It is low enough so as to be considered ``late Universe'' but high enough so that local inhomogeneities are not supposed to have any impact via the so-called cosmic variance on $H_0$~\cite{Camarena:2018nbr}. In other words, GW observations have the potential to shine light in a definitive way on the tension between early- and late-Universe measurements of $H_0$.

So far, different techniques, not mutually exclusive, have been used, all
exploiting the fact that compact binary coalescences are \emph{standard sirens}
\cite{Schutz:1986gp, Holz:2005df}.
If an electromagnetic counterpart is available, then one can break the intrinsic degeneracy between $H_0$ and the coalescence redshift $z$, and precisely determine the Hubble constant with just a few tens of events~\cite{Chen:2017rfc}.%
\footnote{See also \cite{Dalal:2006qt,Nissanke:2009kt} for the role of gamma-ray bursts
  in conjunction with standard sirens, and e.g.~\cite{Belgacem:2019tbw,Zhang:2019loq,Jin:2020hmc,Jin:2021pcv} for measure
of cosmic expansion history by using additional probes than standard sirens.}
The first, and so far unique, of these standard sirens was GW170817 and provided alone a 14\% measurement of $H_0~$\cite{Abbott:2017xzu}.
On the other hand, most of the observed binary coalescences do not
have electromagnetic counterparts and the redshifts of galaxies in the angular position of the coalescence, inferred from galaxy catalogs, can be used to break the $H_0$-$z$ degeneracy~(see \cite{Schutz:1986gp} and the recent \cite{Diaz:2021pem}).
The first of these \textit{dark sirens} was GW170814~\cite{Soares-Santos:2019irc}. Although not yet constraining, given the rapidly increasing number of detections, one expects percentage level constraints after 50 events~\cite{DelPozzo:2011yh}, if catalogs are complete enough (see also \cite{Zhu:2021aat}).
Confining oneself to the binary neutron star case, observation of tidal
  effects can break the gravitational mass-redshift degeneracy, enabling
  the reconstruction of the Hubble relations without electromagnetic
  counterparts \cite{Messenger:2011gi}.
Alternatively, one can exploit the spatial clustering scale between galaxies
and gravitational wave sources, as proposed by
\cite{Mukherjee:2020hyn,Mukherjee:2020mha}:
this method is expected to produce accurate and precise measurements of the
expansion history of the Universe.

Finally, another intriguing method uses the expected gap in 
the black hole mass function due to the pair-instability supernovae
\cite{Heger:2002by}.
Features in the mass distribution break indeed the mass-redshift degeneracy intrinsic to GW observations, so that it is possible to measure $H_0$ without electromagnetic counterparts or host galaxy catalogs~\cite{Farr:2019twy,Ezquiaga:2020tns,You:2020wju,Mastrogiovanni:2021wsd}.

Here, improving on the idea presented in \cite{Ding:2018zrk}, we propose an alternative method to measure the Hubble constant.
This technique uses all observed binary black hole coalescences,
which represent the quasi totality of the events: the $H_0$-$z$ degeneracy of these \textit{black sirens} is broken via the expected (parameter-dependent) redshift
distribution of coalescences.%
\footnote{See also \cite{Ye:2021klk} for a similar idea using binary neutron stars only.}
As we will argue, instead of using galaxy catalogs, unavailable or incomplete
for most events, one can exploit the prior distribution of the coalescence redshift, {suitably} convolved with the instrumental sensitivity of the detectors.
In particular, our method is expected to outperform methods that rely on galaxy catalogs in the limit
of many observations [$O(10^4)$] with poor localization at $z\sim 1-2$.
Therefore, it could be tested with coalescences observed by second generation detectors
during their future runs and it should definitely be efficient
with third generation detectors.

This paper is organized as follows: Sec.~\ref{method} presents the method, whose limiting cases are treated analytically and discussed in Sec.~\ref{limits}. The forecasted results relative to third generation detectors are presented in Sec.~\ref{projection}. We conclude in Sec.~\ref{conclusions}.

\section{Method}\label{method}

Throughout this paper we will adopt the standard model of cosmology, according to which the Universe is spatially flat and has an energy content made of vacuum energy (the cosmological constant $\Lambda$) and pressureless matter (mostly cold dark matter, CDM).
The low-redshift background evolution of the flat $\Lambda$CDM model is  completely specified by the values of the Hubble constant $H_0$ and of the matter density parameter $\Om$. In particular, in our model, the luminosity distance is related to the redshift via:
\begin{align} \label{dL}
d^{(t)}_L(z) &= \frac{c\,(1+z)}{H_0} \int_0^z  \frac{\rd \bar z}{E(\bar z)} \,, \\
E(z) &\equiv \frac{H(z)}{H_0}= \sqrt{\Om\,(1+z)^3 + 1-\Om} \,,
\end{align}
with the comoving distance $d_C^{(t)}=d_L^{(t)}/(1+z)$, with the index $(t)$
standing for ``theoretical''.

Let us now consider one coalescence event. GW detections measure the luminosity distance $d_L$ so that one can build the posterior distribution $f$ of the cosmological parameters and binary black hole (BBH) redshift as follows:
\begin{align}
f(H_0, \Om, z | d_L) =  \frac{ p_{\rm tot}(H_0, \Om, z)  \L(d_L, | H_0, \Om, z)}{\E}  ,  
\end{align}
where here the evidence $\E$ is just a normalization constant.
We will now discuss the prior $p_{\rm tot}$ and the likelihood $\L$.

\subsection{Prior} \label{sec:prior}

Using the product rule, the prior can be written as:
\begin{align} 
p_{\rm tot}(H_0, \Om, z)=p(H_0) \, p(\Om) \,p(z|H_0, \Om) \,.  \label{prior}
\end{align}
We assumed that $\Om$ and $H_0$ are independent because for the former we use an informative prior from Supernovae Ia, which is  independent from $H_0$.
We adopt the almost Gaussian prior from the Pantheon dataset \cite{Scolnic:2017caz}:
\begin{equation}
p(\Om) \propto \exp \left[ - \frac{\left (\Om - \Om^{(p)} \right)^2}{2 \sigma^2_{m,p}} \right ] \,,
\end{equation}
where $\Om^{(p)}=0.298$ and  $\sigma_{m,p}=0.022$.

Regarding $p(H_0)$, as we aim at measuring the Hubble constant with black sirens, we adopt a flat broad prior:
\begin{equation}
p(H_0) \propto \left\lbrace
\begin{array}{cl}
\text{const} & \text{if } H_0 \in [20,140]\text{ km } \text{s}^{-1} \text{Mpc}^{-1} \\
0 & \text{otherwise}
\end{array}
\right. ,
\end{equation}
which is the same prior adopted by \cite{Soares-Santos:2019irc}.

The prior on the observed coalescence redshift
$p(z|H_0, \Om)$ is the nontrivial piece of information necessary to extract information on $H_0$ from gravitational wave observations.
The standard dark-siren approach  estimates the redshift prior via a galaxy catalog that covers the sky localization of the event \cite{Soares-Santos:2019irc,Fishbach:2018gjp,Abbott:2019yzh}.
This approach has the advantage of correlating the coalescence to the actual nearby galaxies and, in particular, to their large-scale structure of voids, filaments and clusters.
However, the galaxy catalog may be incomplete or unavailable.
The idea at the base of our black-siren method is to estimate $p(z|H_0, \Om)$ theoretically.
More precisely, in the present paper we will obtain the redshift prior via an analytical estimation of the star-formation rate, convolved with a suitable
star formation to binary coalescence delay, while we leave for future work
the use of synthetic galaxy catalogs from state-of-the-art hydrodynamical  simulations.

We model the redshift prior via two contributions:
\begin{align} 
  p(z|H_0, \Om) = A(H_0, \Om) \, R_m(z) \, f_C \! \left(d^{(t)}_C(z) \right ) \,,
  \label{coa-prior}
\end{align}
which we now explain in detail. In the previous equation $A$ is a normalization constant which may depend on all the parameters but $z$. 

\subsubsection{Merger rate}

The first contribution $R_m(z)$ is the rate number ($N_m$) density of
mergers in the detector frame (number of mergers per detector time per
redshift) which will be expressed via:
\begin{equation} \label{ptau}
 R^{(\tau)}_m(z) \equiv \frac{\rd N^{(\tau)}_m}{\rd t_d \rd z} \,,
\end{equation}
where we omit the inconsequential normalization constant and the hyper parameter
$\tau$  is discussed below.
Following  \cite{Vitale:2018yhm,Mendonca:2019yfo}, we model $R_m$ via the
total merger rate per comoving volume in the source frame ${\cal R}^{(\tau)}_m\equiv \frac{\rd N_m^{(\tau)}}{\rd V \rd t_s }$:
\begin{equation}
  \label{eq:merger_rate}
 R^{(\tau)}_m(z) = \frac 1{1+z}\frac{\rd V}{\rd z}{\cal R}^{(\tau)}_m(z) \,,
\end{equation}
where the $1 + z$ term in the denominator arises from converting source-frame time $t_s$ to detector-frame time $t_d$, and  $\rd V/\rd z$ is the cosmology-dependent comoving volume element per unit redshift interval:
\begin{align} \label{vol}
\frac{\rd V}{\rd z} &= \frac{4 \pi}{H(z)} \,  \frac{c\, d_L^2(z)}{(1+z)^{2}} = \frac{4 \pi}{E(z)}  \pa{\frac c{H_0}}^3 \paq{\int_0^z  \frac{\rd \bar z}{E(\bar z)} }^2 .
\end{align}
Then, we model ${\cal R}^{(\tau)}_m$ via a delayed volumetric BBH formation rate ${\cal R}_f$.
Specifically, we account for the stochastic delay between star formation
and BBH merger via a Poissonian distribution of characteristic delay $\tau$:
\be
   {\cal R}^{(\tau)}_m(z)=\frac 1\tau\int_z^\infty \rd z_{f}\frac{\rd t}{\rd z_{f}}{\cal R}_f(z_{f})
   \exp\paq{-\frac{t(z_{f})-t(z)}\tau}\,, \label{poisson}
\ee
where
\begin{align}
  t(z)\equiv\frac 1{H_0}\int_0^z \frac{\rd\bar z}{(1+\bar z)E(\bar z)}
\end{align}
is the time spent between redshift $z$ and the present epoch.
Note that ${\cal R}^{(\tau)}_m$, apart from the normalization, depends on $\tau$
only via the dimensionless combination $H_0 \tau$.

Finally, we assume that the BBH volumetric formation rate is proportional to the star formation rate density $\psi(z)$ at the same redshift:
\begin{equation}
  \label{eq:formation}
{\cal R}_f(z_{f}) \equiv \frac{\rd N_f}{\rd V \rd t_f } \propto \psi(z_f)  \,.
\end{equation}
In other words we are not considering the time between star formation and BBH formation, which should be negligible given the time scale of BBH coalescence.
We adopt the measured star formation rate from \cite{Madau:2014bja}:
\begin{equation}
\label{eq:sfrDM}
\psi_{\rm MD14}(z) =0.015 \frac{(1+z)^{2.7}}{1+ \left( \frac{1+z}{C} \right)^{5.6}} \, M_\odot \text{ yr}^{-1} \text{Mpc}^{-3} \,,
\end{equation}
with $C=2.9$.
The merger rate obtained using Eq.~(\ref{eq:sfrDM}) in Eq.~(\ref{eq:formation})
may not correspond to the one realized in nature.
We do not account here for the fact that only a fraction of stars ends up in
black holes. Moreover we neglect that both merger rate and time delay
distribution may depend on binary intrinsic properties, like component masses
and spins.
Such dependences can be modeled by including additional hyperparameters to
the proposed merger rate and eventually marginalizing over them, at the
cost of degrading the precision of the recovery of cosmological parameters.
However,  we will neglect these details for the moment to show in principle
the power of the method, and in the Appendix we show that
the addition of another hyperparameter
can absorb the effect of our ignorance of the underlying merger rate, and
still produce an unbiased  determination of the Hubble constant, at the price
of moderately degrading the precision of parameter estimation.
See \cite{LIGOScientific:2021aug} for a recent application
of jointly fitting the cosmological parameters and the source population
properties of binary black holes.

As already mentioned, the characteristic delay $\tau$ is a hyperparameter of
the redshift prior. We adopt a flat  hyperprior:
\begin{equation}
p(\tau) \propto \left\lbrace
\begin{array}{cl}
\text{const} & \text{if } \tau \in [100\ {\rm Myr},t_0(H_0, \Om)] \\  
0 & \text{otherwise}
\end{array}
\right. ,
\end{equation}
where $t_0$ is the age of the Universe (since we observe the coalescence it must be $\tau<t_0$). 
One can then consider the following compound distribution as the coalescence prior:
\begin{align}
  \label{eq:rm}
R_m(z) = \int_0^\infty \rd \tau \, p(\tau)  \, R^{(\tau)}_m(z) \,.
\end{align}
Note that, numerically, it is equivalent to include $\tau$ as a nuisance parameter with prior $p(\tau)$. We will adopt this point of view when considering
a generic number $n$ of events.

\subsubsection{Detector sensitivity} \label{detector}

\begin{figure}
  \begin{center}
    \includegraphics[width=\columnwidth]{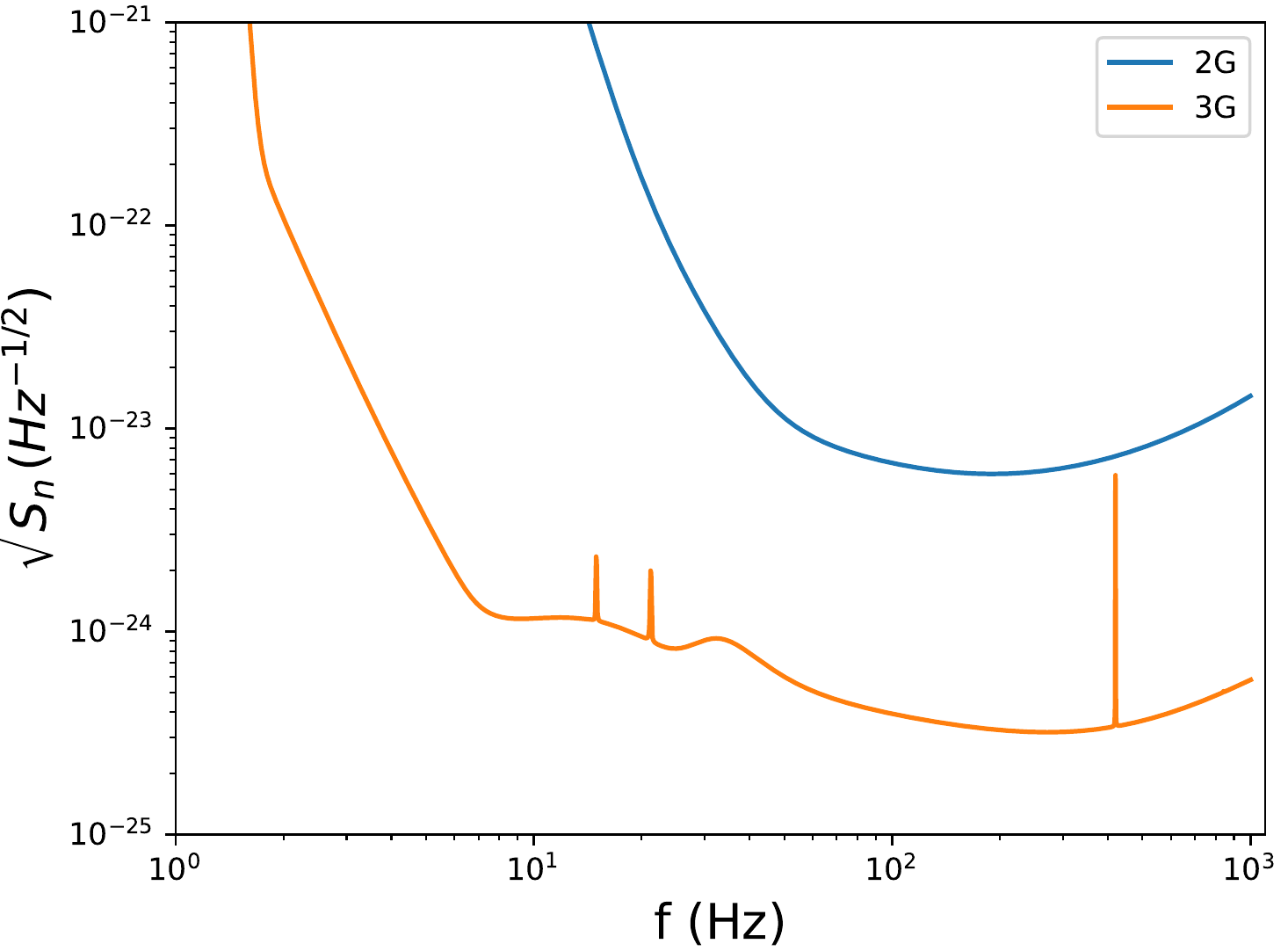}
  \end{center}
  \caption{Second and third generation  spectral noise densities.
    The 2G noise curve has been obtained by a fit to the LIGO Hanford O3 data around the event GW190814
    based on the ``Zero Detuning High Power'' spectral noise density implemented in LALSuite \cite{lalsuite}.
    For the 3G curve we adopted the noise spectral density ``D'' from \cite{Hall:2019xmm}.}
  \label{fig:noise}
\end{figure}

The last piece in Eq.~\eqref{coa-prior}, $f_C(d_C^{(t)})$, models the LIGO-Virgo detector
sensitivity on the luminosity distance: obviously more distant sources
are less likely to be detected than nearer ones.
Indeed, coalescences are observed if a signal-to-noise ratio ($SNR$) larger than 8 is achieved.
The $SNR$ is computed by comparing the $f$-domain waveform $\tilde h(f)$ with the detector noise $S_n$:
\begin{align}
SNR &=2\paq{\int_0^\infty df \, \frac{|\tilde h(f)|^2}{S_n(f)}}^{1/2}\,, \label{snr} \\
\tilde h(f)&=F_+ \tilde h_+(f)+F_\times\tilde h_\times(f) \,,
\end{align}
where the pattern functions $F_{+,\times}$ are function of the two angles locating the source in the
sky ($\alpha,\delta$) and the polarization angle $\psi$, and the GW
polarizations $\tilde h_{+,\times}$ are given at leading order (quadrupole formula) by:%
\footnote{Note that the interference term between $\tilde h_+$ and $\tilde h_\times$ vanishes in the $SNR$ integral. Analytic expressions (\ref{eq:h+},\ref{eq:hx}) are shown for illustration and are valid only for the inspiral phase of the coalescence.}
\begin{align}
\label{eq:h+}
  \tilde h_+&=\ds\pa{\frac 5{24}}^{1/2}\frac{\pi^{-2/3}}{d^{(t)}_L} \Mz^{5/6}
  f^{-7/6}\pa{\frac{1+\cos^2\iota}2}^2e^{i\phi(f)} ,\\
  \label{eq:hx}
  \tilde h_\times&=\ds\pa{\frac 5{24}}^{1/2}\frac{\pi^{-2/3}}{d^{(t)}_L} \Mz^{5/6}
  f^{-7/6}\cos\iota\ e^{i\phi(f)+i\pi/2} \,,
\end{align}
which depend on the luminosity distance, the orientation $\iota$ and the redshifted chirp mass $\Mz \equiv M_c (1+z)$.
The chirp mass is defined by $M_c\equiv \eta^{3/5} M$, where $M\equiv m_1+m_2$, $\eta \equiv m_1m_2/M^2$, and $m_i$ are the individual constituent masses. The angle $\iota$ gives the relative orientation between the binary orbital plane and the observation direction.
Fig.~\ref{fig:noise} shows the square root of the noise spectral density $\sqrt{S_n}$ used to estimate the $SNR$ for second (2G) and third (3G) generation detectors.

\begin{table}
  \caption{Parameter space that is uniformly explored (except for masses) to
    sample the $SNR$ of Eq.~\eqref{snr}.
    For the individual masses the distribution adopted is a broken power law
    $\propto m_i^{-1.5}$ $(m_i^{-5})$ for $m_i<40M_\odot$ $(40<m_i/M_\odot <80)$ for
    solar mass black holes and a log prior for intermediate masses
    $120 < M/M_\odot<10^4$.}
\label{tab:mc}
\begin{tabular}{lcc}
\hline
\hline
Parameter & Quantity   & Interval \\
\hline
\hline
Comoving distance & $d_C^{(t)}/$Mpc  & $[100, 1.2\cdot 10^4]$ \\
Individual mass & $m_i/M_\odot$ &  $[1.2,10^4]$\\
Mass ratio & $q=m_2/m_1$ & $>10^{-3}$\\
Binary orientation & $\cos\iota$ & $[-1,1]$ \\
Polarization & $\psi$ & $[0,2 \pi]$ \\
Right ascension & $\alpha$ & $[0, 2 \pi]$\\
Declination & $\delta$ & $[0, \pi]$\\
\hline
\hline
\end{tabular}
\end{table}

To relate the astrophysical to the \emph{detected} merger rate one needs
to take into account selection effect, i.e.~to estimate how likely it is to detect a source located at a given distance
from the observatory, which is obtained by averaging over the source parameters
to get the average distribution of detections as a function of distance.
The requirement for  detection is that the signal has $SNR\geq 8$,
and averaging is performed over masses and angles as reported in Table~\ref{tab:mc}.

The astrophysical mass distribution of stellar-mass black holes can be inferred from LIGO/Virgo O1, O2, O3a data
as described in \cite{LIGOScientific:2018jsj,LIGOScientific:2020kqk}. This is relevant for 2G detectors as they are sensitive to binaries with total mass up to $\sim O(100 M_\odot)$.
We can assume that the mass of the heavier binary component
is distributed according to a broken power law with exponents $\alpha_1=-1.5$
and $\alpha_2=-5$ for masses between $5$ and $60$ $M_\odot$, with
the slope change occurring at $m_{\rm break}=40 M_\odot$. The mass ratio $q$ is assumed
to be distributed according to $p(q)\propto q^{-1}$ with $0.1\leq q\leq 1$, with a lower
cutoff on the lighter mass assuming $1.2M_\odot <m_2$.
Third generation detectors will be also sensitive to intermediate-mass black holes with
$m_i\gtrsim 10^2 M_\odot$.
As their distribution is completely unknown, we have assumed a mass gap from
$80$ to $120 M_\odot$ due to {pair-instability supernovae}
\cite{Heger:2002by} and an uninformative $\rd\log$ prior up to $m_i<10^4M_\odot$.
In the same spirit of the 2G case, that is to use a concrete example to test the method, we assume the distribution of the primary mass to be $\propto 1/m_1$ for
$120 \leq m_1/M_\odot\leq 10^4$ and for the mass ratio in this region the prior $p(q)\propto q^{1/2}$.

It is important to stress that stellar- and intermediate-mass black hole population properties are
not precisely known and that here we wish to use indicative values for the underlying
population to test the efficiency of our method in a realistic case.
Moreover, the black hole mass function is only used to evaluate the reach
  of the detector. Besides this detail, its information is \emph{not} folded
  into the likelihood to determine cosmological parameters.
As we will show, the method proposed here can lead to interesting constraints on $H_0$
 only for a large number of detections $\gtrsim O(10^3)$.
Hence, we can safely assume that once accumulating so many detections, the
population properties of the sources will be known with great accuracy.
The use of a different underlying astrophysical mass distribution will impact both
the simulated signals and the priors entering the determination of the $H_0$
posterior probability distribution, leaving basically unaltered the predictive power of
the method.

\begin{figure}
  \begin{center}
    \includegraphics[width=\columnwidth]{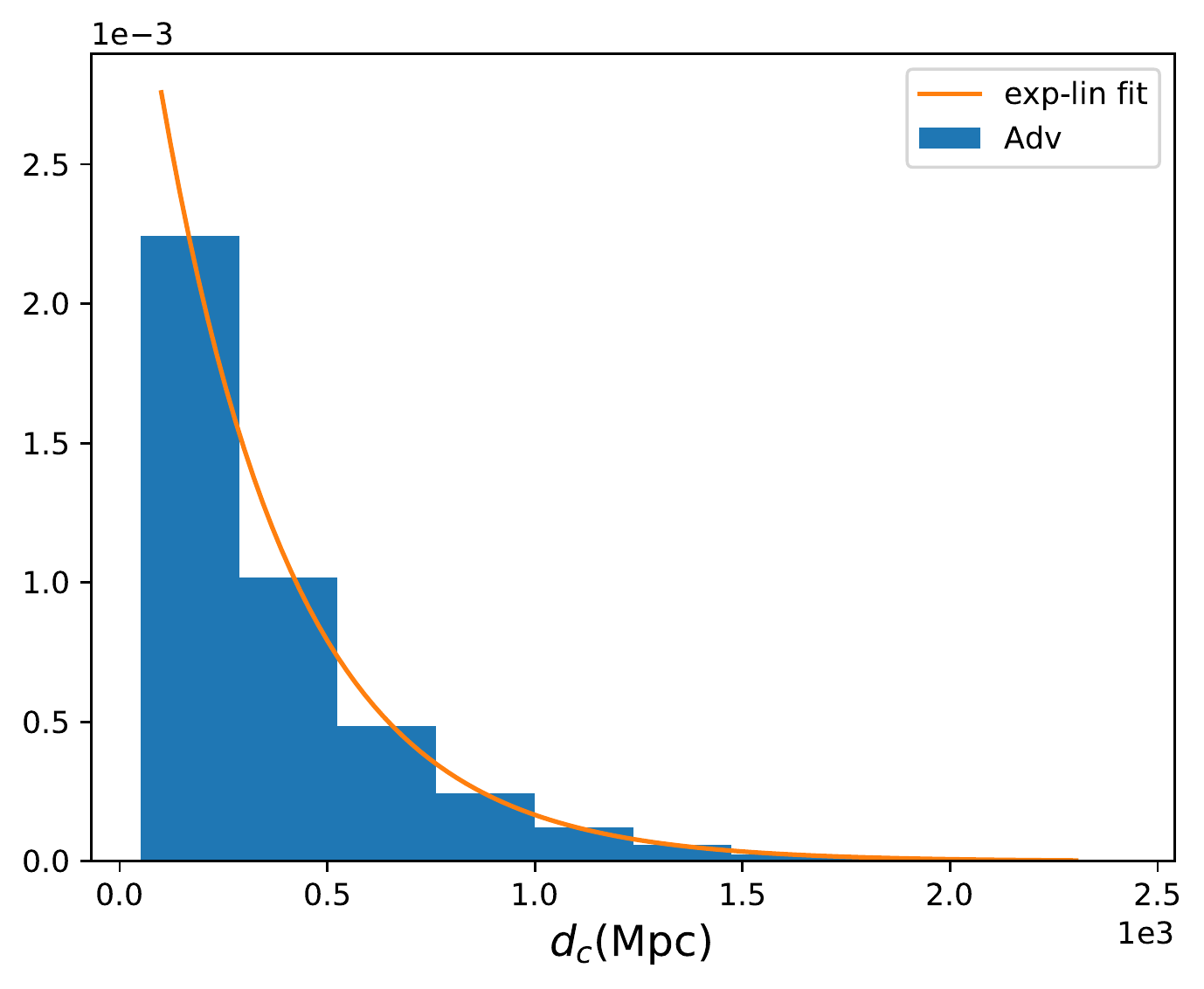}
    \includegraphics[width=\columnwidth]{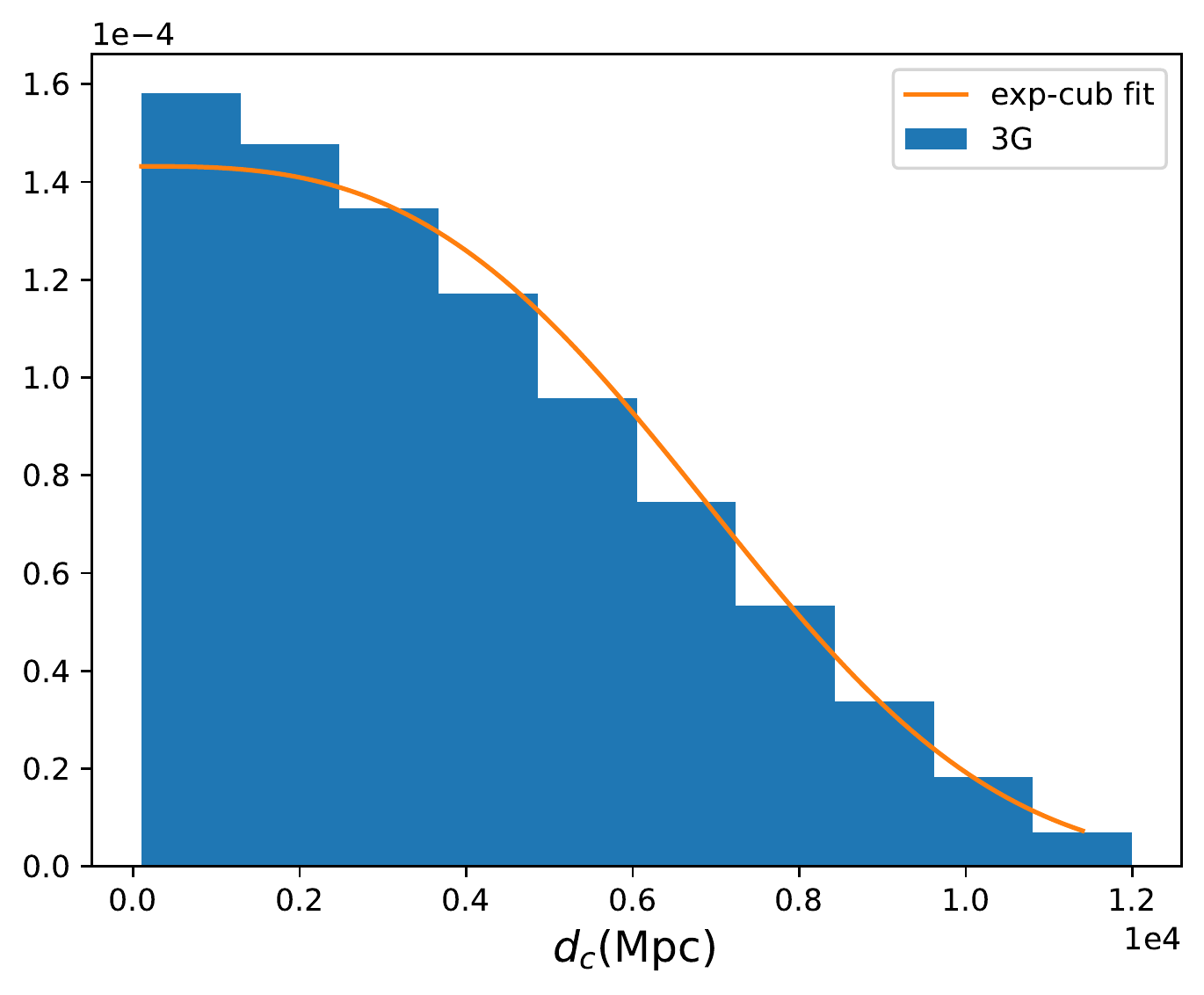}
  \end{center}
  \caption{Distribution of the comoving distance $d_C$ for the events satisfying
    $SNR>8$, see Eq.~\eqref{snr}, averaged over masses and orientations for 2G
    and 3G detectors. The curve shows the fit of Eq.~\eqref{eq:fitdc} to the
    tail of the distribution. Note the different scale for both axis in the two figures.}
  \label{fig:distdc}
\end{figure}

We use the waveform approximant known as \emph{IMRPhenomD} \cite{Husa:2015iqa,Khan:2015jqa}, describing the entire coalescence, for spinless sources
generated via LALSuite \cite{lalsuite}, and noise as in Fig.~\ref{fig:noise}, representative
of second and third generation ground-based GW detectors.
After imposing $SNR>8$ and averaging over all parameters but $d_C^{(t)}$, we obtain the distributions $f(d_c^{(t)})$ shown in Fig.~\ref{fig:distdc} whose tail
in the 2G and 3G cases can be modeled according to:
\begin{align}
  \label{eq:fitdc}
  f_C\left(d_C^{(t)}\right ) \propto\left\{
  \begin{array}{lr}
  \exp\left[ -\frac{d^{(t)}_C}{\dccs} \right ] &{\rm 2G}\\
  \ &\\
  \exp\left[ -\pa{\frac{d^{(t)}_C}{\dcct}}^3 \right ] &{\rm 3G}
  \end{array}\right.\,,
\end{align}
where $\dccs= 320$ Mpc and $\dcct= 7.9$ Gpc.

The decay with the comoving distance is qualitative different in the 2G and 3G
cases.
In the 2G case, only sources at moderate redshift are visible, as
increasing the distance increases the denominator in Eqs.~(\ref{eq:h+},\ref{eq:hx}),
thus decreasing the $SNR$.

\begin{figure}
\begin{center} 
  \includegraphics[width=\columnwidth]{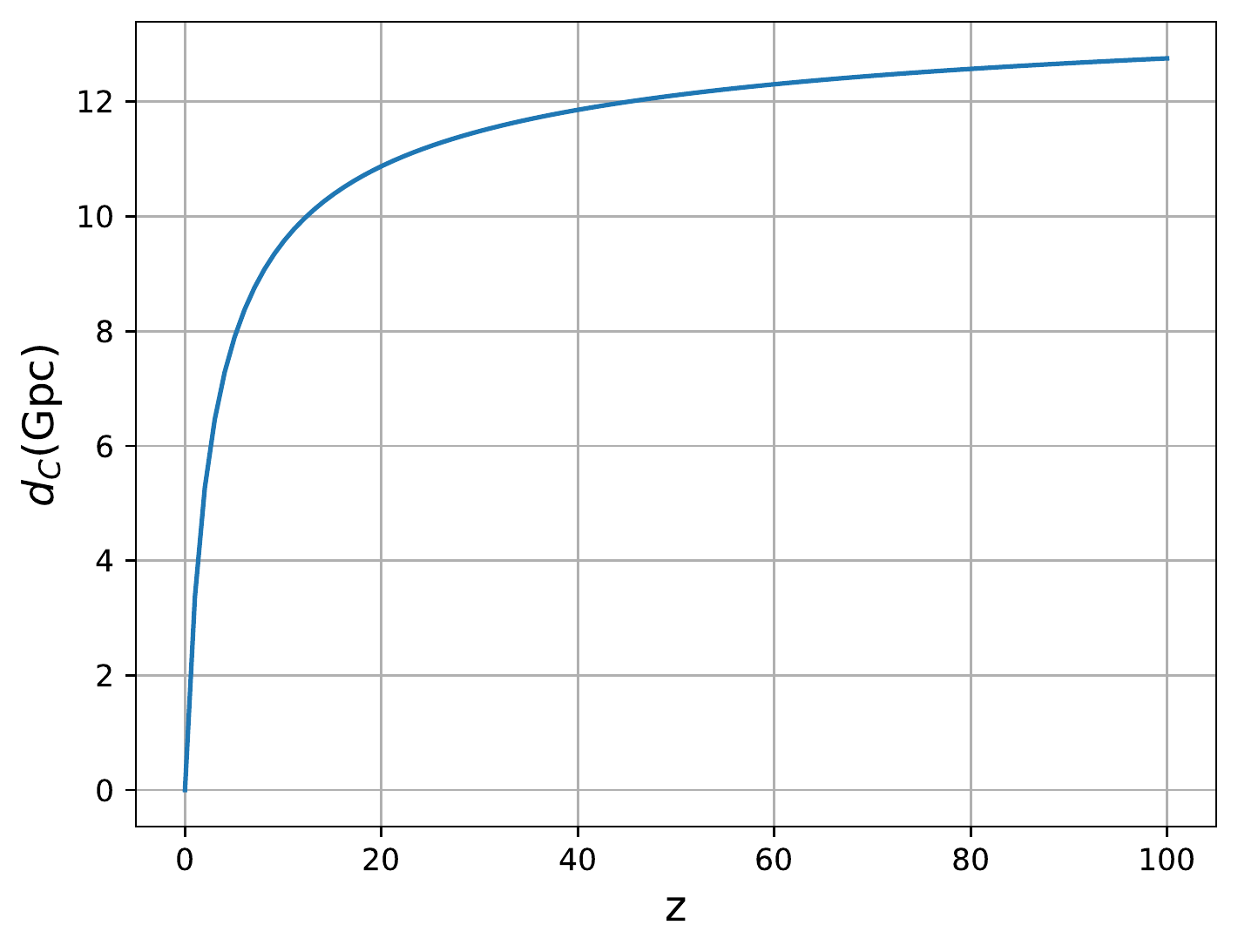}
  \caption{Comoving distance-redshift relationship for standard $\Lambda$CDM model.}
  \label{fig:dcz}
\end{center}
\end{figure}

In the 3G case,  signals with $z>1$ are visible for a wide
range of masses, with the result that the $(1+z)^{5/6}$ dependence at the numerator of Eqs.~(\ref{eq:h+},\ref{eq:hx})
almost cancels the $z$-dependence of $d_C(1+z)$ at the denominator.
As a consequence, the $SNR$ varies with distance approximately according to
$d_C^{-1}$ until the redshift pushes the
signal to low enough frequencies to fall outside the detector's band,
and this happens around $d_C\simeq 12 $ Gpc for a wide range of masses,
as that is the value at which $z$ steeply increases for small variation of $d_C$, see Fig.~\ref{fig:dcz}.

Note that the $SNR$ depends on the redshifted chirp mass $\Mz$ which depends on
redshift.
To obtain the simulations presented in Fig.~\ref{fig:distdc} the redshift is not
varied independently but instead determined from the distance and the fiducial
cosmology ($\Lambda CDM$):
$z^{(\rm fid)}=z(d_C^{(t)}, H_0^{(\rm fid)}, \Om^{(\rm fid)}) \overset{z\rightarrow 0}{\simeq}  H_0^{(\rm fid)} d_C^{(t)}/c$, with $H_0^{(\rm fid)}=69.32 \text{ km s}^{-1} {\rm Mpc}^{-1} $  and $\Om^{(\rm fid)}=\Om^{(p)}$.

\subsection{Likelihood}

In the Gaussian approximation, the likelihood can be written according to:
\begin{align}
\L(d_L, | H_0, \Om, z)  \propto  \exp \left[ - \frac{\big(d_L- d_L^{(t)}(H_0, \Om, z) \big)^2}{2 \sigma_L^2} \right ]\,,
\end{align}
where to lighten notation the dependence of the likelihood on the luminosity distance uncertainty $\sigma_L$ has been suppressed.

\subsection{Posterior for $n$ coalescences}

When combining $n$ coalescences it is convenient to marginalize immediately on the parameters that are specific to a given event so that:
\begin{align}
f(H_0, \Om , \tau|  d_L) &\propto p(H_0) \, p(\Om) \,  p(\tau)  \\
& \times \int   \rd z  \, p(z|H_0, \Om,\tau)  \, \L(d_L, | H_0, \Om, z) \,,\nonumber
\end{align}
where, as discussed earlier, we treated $\tau$ as a nuisance parameter.
The expression above can then be generalized to the case of $n$ detections $ \{d_{L,i} \}$:
\begin{align}
 f(H_0, \Om, \tau &|  \{d_{L,i} \}) \propto p(H_0) \, p(\Om) \, p(\tau)  \label{postn} \\
& \times  \prod_{i=1}^n  \int \rd z_i  \, p(z_i|H_0, \Om,\tau) \, \L(d_{L,i} | H_0, \Om, z_i) .  \nonumber
\end{align}
Numerically, the posterior exploration will be performed on the parameters $H_0, \Om, \tau$.
In other words, for each point $\{ H_0, \Om, \tau \}$ of the parameter space we will estimate the $n$ 1-dimensional integrals of Eq.~\eqref{postn}.
We parametrize here the inevitable uncertainty in the knowledge of the
underlying merger distribution with only one hyperparameter $\tau$,
and we address in the Appendix the issue of the generality of the merger rate
function that we adopt in Eq.~\eqref{eq:rm}.

\section{Limiting cases} \label{limits}

To understand analytically the statistical inference on $H_0$ with black sirens it is useful to consider the following limiting cases.

\subsection{Low redshift}

It is interesting to take the limit $z \rightarrow 0$ in Eq.~\eqref{coa-prior}. First, one has that $\frac 1{1+z}\frac{\rd V}{\rd z} \sim z^2 \sim d_L^2$.
Second, from Eq.~\eqref{eq:sfrDM} it follows that ${\cal R}_f(z) \sim $ constant so that, from Eq.~\eqref{poisson}, one finds that ${\cal R}^{(\tau)}_m(z)\sim $ constant. One then finds from Eq.~\eqref{coa-prior} that:
\begin{equation} \label{zlimit}
p(z|H_0, \Om)  \overset{z\rightarrow 0}{\propto}  d_L^2 \,.
\end{equation}
In other words, the prior cannot break the $H_0$-$z$ degeneracy as it depends just on $d_L$, which is the quantity measured by GW observations.
Equivalently, the information that is able to break the $H_0$-$z$ degeneracy comes from a nontrivial ${\cal R}_f$.

\subsection{Negligible luminosity distance error}

Next, we can take the limit $\sigma_L/d_L \rightarrow 0$ in Eq.~\eqref{postn}:
\begin{align}
& f(H_0, \Om, \tau |  \{d_{L,i} \}) \propto p(H_0) \, p(\Om) \, p(\tau)  \nonumber \\
& \times  \prod_{i=1}^n  \int \! \rd z_i  \, A \, R^{(\tau)}_m(z_i) \, e^{-{d^{(t)}_L(z_i)}/{ \dc}}    \delta(d_{L,i} \!-\! d_L^{(t)}(H_0, \Om, z_i) )  \nonumber \\
 &= p(H_0)  p(\Om)  p(\tau)  \prod_{i=1}^n  \frac{A \, R^{(\tau)}_m(z_i, H_0 \tau, \Om) \, e^{-{d_{L,i}}/{ \dc}}}{\left| \frac{\partial d_L^{(t)}}{\partial z^{(t)}}(H_0, \Om, z_i)\right |} , \label{postn0}
\end{align}
where we used the properties of the Dirac delta function and $z_i=z^{(t)}(d_{L,i}, H_0, \Om)$ is the theoretical redshift associated with $d_{L,i}$ given $H_0$ and $\Om$, and assuming $f_C=e^{d_L/d_L^{({\rm cut})}}$.
We see that, in this limit, the detector sensitivity $f_C$ does not contain cosmological information.

\subsection{Infinite number of observations}

Statistical inference with black sirens suffers from two sources of uncertainties.
The first is due to the uncertainty $\sigma_{L}$ on the measurement of the luminosity distance.
The second comes from having a finite sample $n$ of observations. Indeed, we are constraining parameters to recover the actual distribution of coalescence
redshifts.

From Eq.~\eqref{postn0} it is easy to see how a fiducial model is recovered in the limit of infinite observations. Assuming flat priors on $H_0$, $\Om$ and $\tau$:
\begin{align}
f(H_0, \Om, \tau |  \{d_{L,i} \}) &\propto  \prod_{i}^n f (d_{L,i} | H_0, \Om, \tau) \,,
\end{align}
where $f(d^{(t)}_{L} | H_0, \Om, \tau)$ is the theoretical distribution in the luminosity distance given the theoretical model (the Jacobian is absorbed by the change of variable).
From the previous equation one sees that in the limit $n \rightarrow \infty$ the values of $H_0$, $\Om$ and $\tau$ that maximize the posterior are the ones that were used to produce the measurements $\{d_{L,i} \}$.

\subsection{Toy example}

\begin{figure}
\begin{center} 
  \includegraphics[width=\columnwidth]{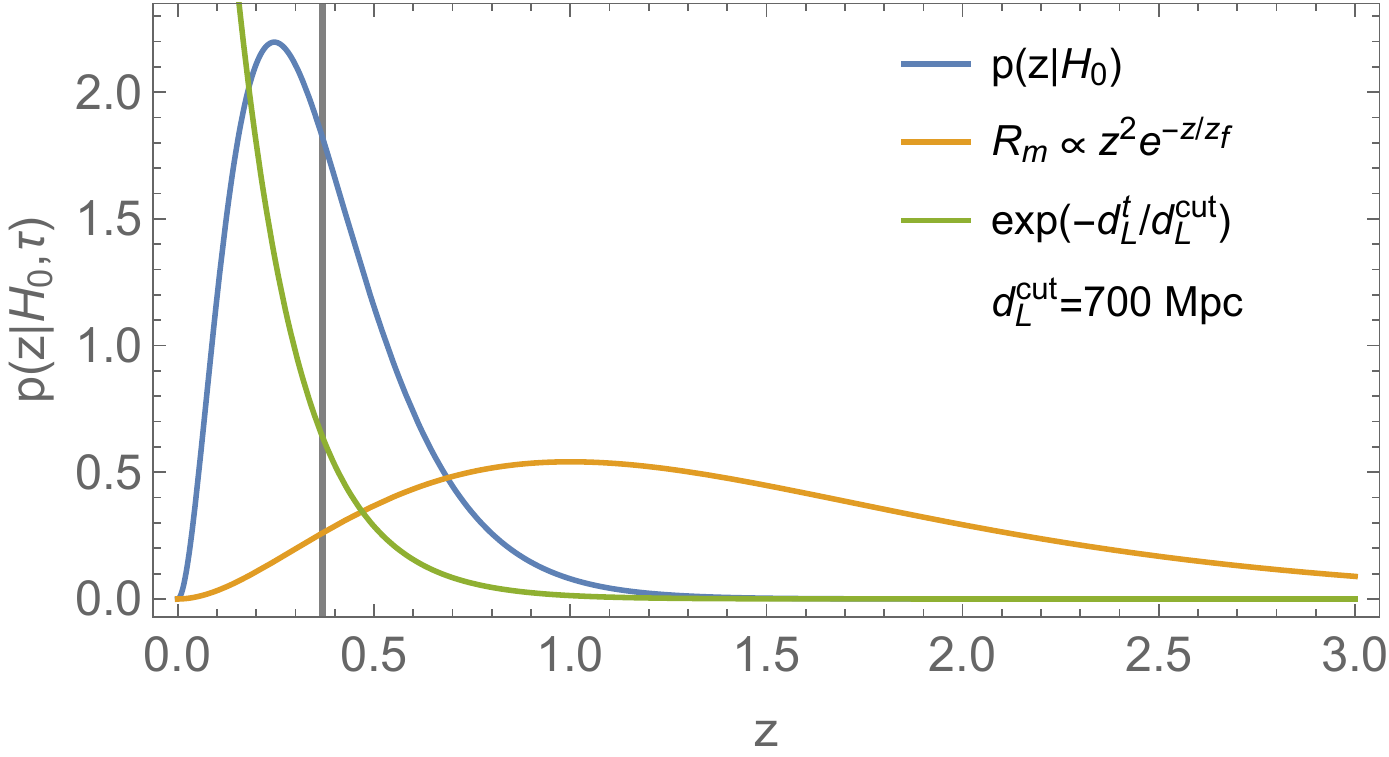}
  \includegraphics[width=\columnwidth]{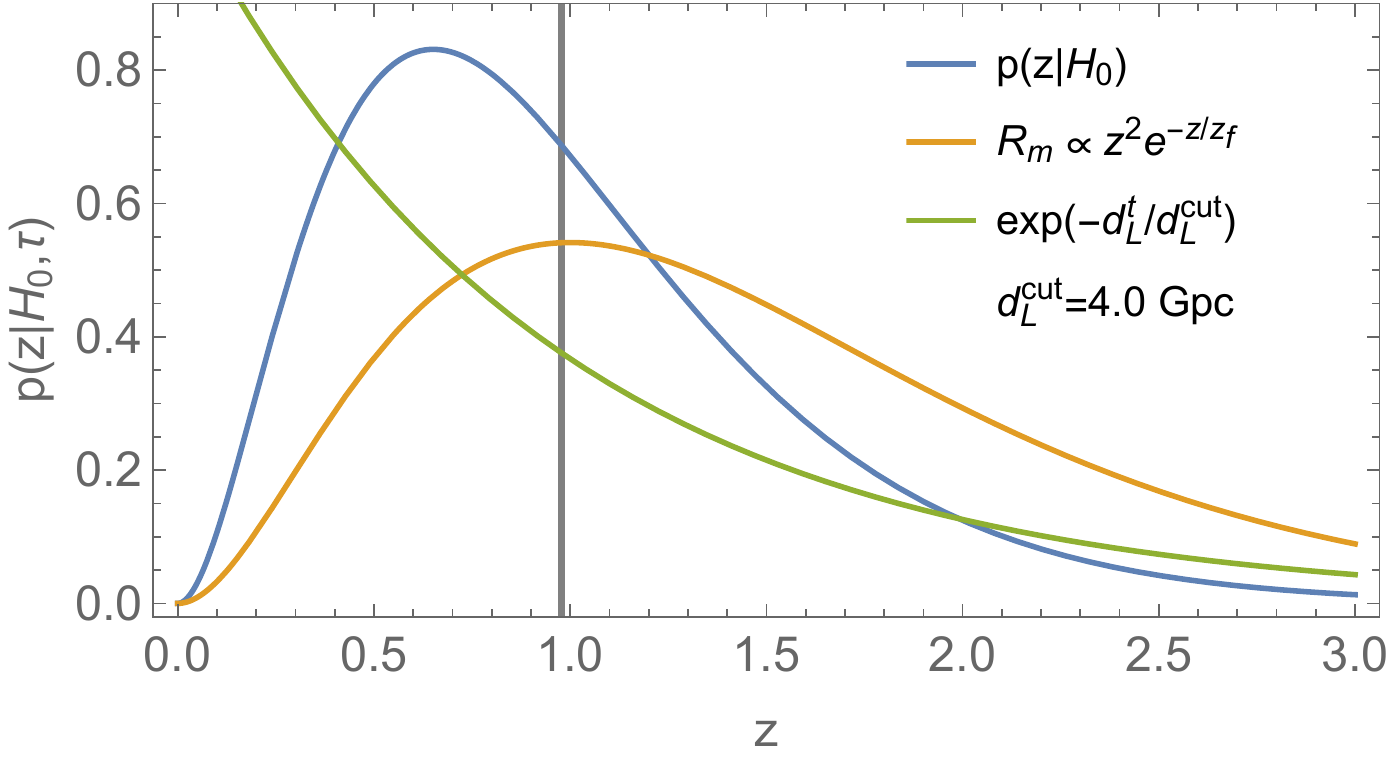}
\end{center}
\caption{Toy redshift prior of Eq.~\eqref{eq:toy-prior} for $H_0^{(\rm fid)} =70 \text{ km s}^{-1} {\rm Mpc}^{-1} $, $z_f =0.5$ and $\dc =700$ Mpc (top) or $\dc =4.0$ Gpc (bottom). The vertical lines mark the mean redshifts.}
\label{fig:toy-prior}
\end{figure}

To further simplify the analysis we consider the following redshift prior:
\begin{align}
p(z|H_0) = \left ( \frac{c}{H_0 \dc}+ \frac{1}{z_f} \right)^3 \, \frac{z^2}{2} \, e^{-\frac{z}{z_f}}  \,  
 \exp\left( \frac{-c \, z}{H_0 \dc} \right ) ,  \label{eq:toy-prior}
\end{align}
where we adopted the approximation $d_C^{(t)}\simeq d_L^{(t)} \simeq c\, z/H_0$, so that we can drop the (anyway weak) dependence on $\Om$.
Eq.~(\ref{eq:toy-prior}) represents a normalized, reasonable toy model where
the factor $z^2 e^{-z/z_f}$ intends to reproduce
the astrophysical merger distributions and a detector sensitivity exponentially
decaying with redshift has been assumed.
Fig.~\ref{fig:toy-prior} shows this prior for two values of the detector luminosity cut  $\dc$.
The vertical lines mark the mean redshifts $\bar z=3z_f z_c /(z_f+z_c)$, where $z_c (H_0 ) =H_0 \dc /c$.

Taking again the limit $\sigma_L/d_L \rightarrow 0$, the posterior becomes:
\begin{align}
\ln f(H_0 |  \{d_{L,i} \}) = &3 n \ln \left(  \frac{H_0 \dc}{c\, z_f} +1  \right ) -n \frac{H_0 \bar d_L}{c \, z_f} \label{postoy} \\
 \overset{\text{average}}{=} &
 3 n \left [ \ln \left( 1 + \frac{H_0 \dc}{c \, z_f} \right ) - \frac{ H_0 \dc}{H_0^{(\rm fid)} \dc + c\, z_f}  \right ] ,\nonumber 
\end{align}
where $\bar d_L\equiv \frac 1n\sum_i d_{L,i}$, we omitted additive constants and in the last equation we used:
\begin{align}
\bar d_L =  \bar z \frac{c}{H_0^{(\rm fid)}} = \frac{3z_f \, z_c(H_0^{(\rm fid)}) }{z_f+z_c(H_0^{(\rm fid)})} \frac{c}{H_0^{(\rm fid)}}\,.
\end{align}
The posterior maximum (best fit) is found by solving $\partial \ln f / \partial H_0 =0$, which gives $H_{0,{\rm bf}}=H_0^{(\rm fid)}$, that is, the fiducial value of the Hubble constant is recovered in the limit of infinite (infinitely precise) measurements.

Finally, we can compute the Fisher matrix, which, in this case, is just a number:
\begin{align}
F=-\left. \frac{\partial^2\ln f(H_0|\{d_{L,i} \})}{\partial H_0^2} \right|_{H_0^{(\rm fid)} }\,,
\end{align}
so that:
\begin{align}
\frac{\sigma_{H_0}}{H_0} =\frac{F^{-1/2}}{H_0} = \frac{1+z_f/z_c (H_0^{(\rm fid)})}{\sqrt{3 n}} \,,
\end{align}
which  depends on $z_c  =H_0^{(\rm fid)} \dc /c$.

\begin{figure}
\begin{center} 
\includegraphics[width=\columnwidth]{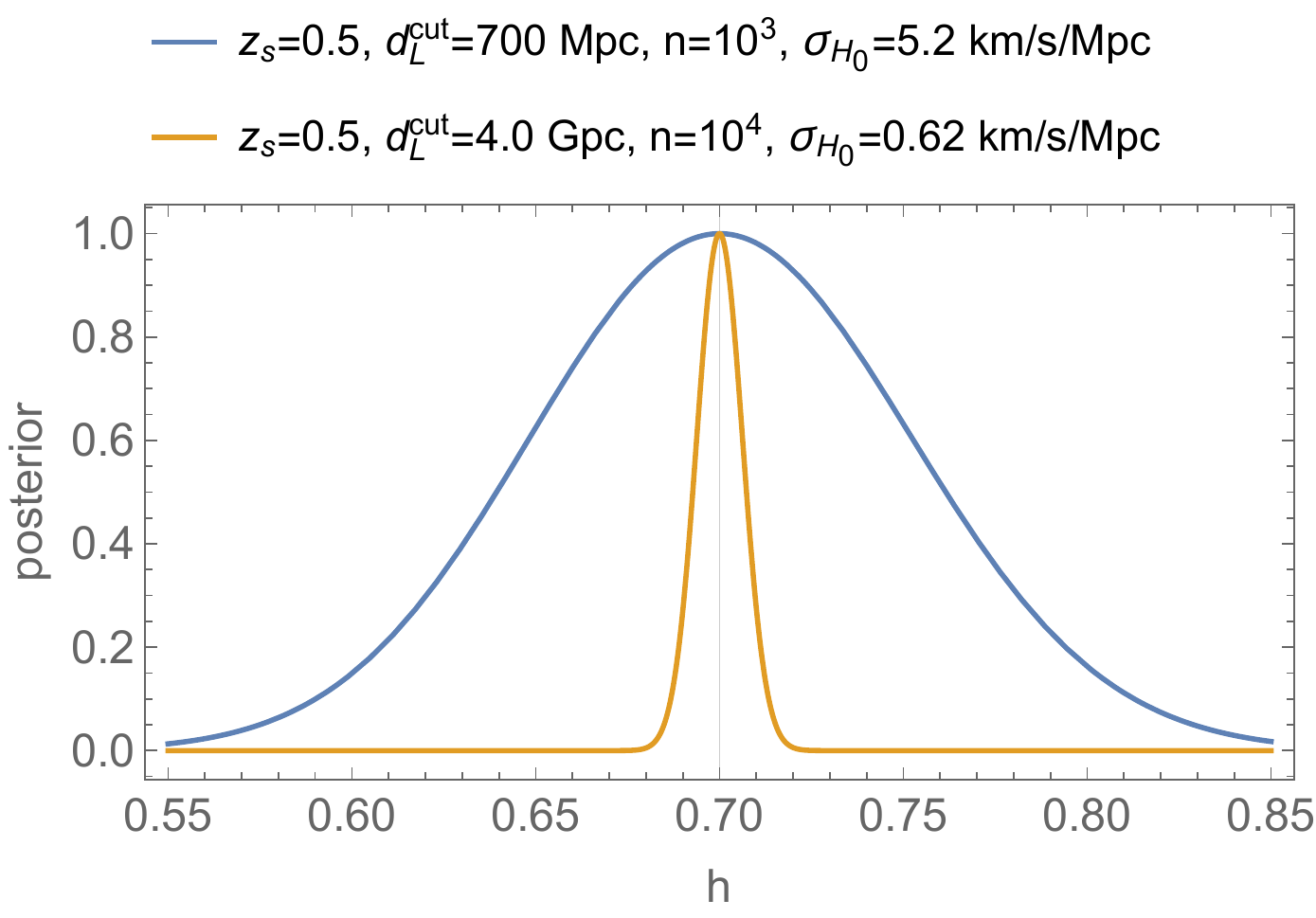}
\end{center}
\caption{Forecasted constraints on $H_0$ relative to the toy model of Eq.~\eqref{postoy}  for a second generation (blue line) and third generation (orange line) detector.}
\label{fig:toy-post}
\end{figure}

Fig.~\ref{fig:toy-post} shows the forecasted constraints relative to the toy model of Eq.~\eqref{postoy} for a second generation (blue line) and third generation (orange line) detector.
This result does not take into account the degeneracy of $H_0$ with $\Om$ and $\tau$. In the next Section we will discuss a realistic forecast.

\section{Realistic forecast} \label{projection}

\begin{figure}
  \begin{center}
    \includegraphics[width=.9\columnwidth]{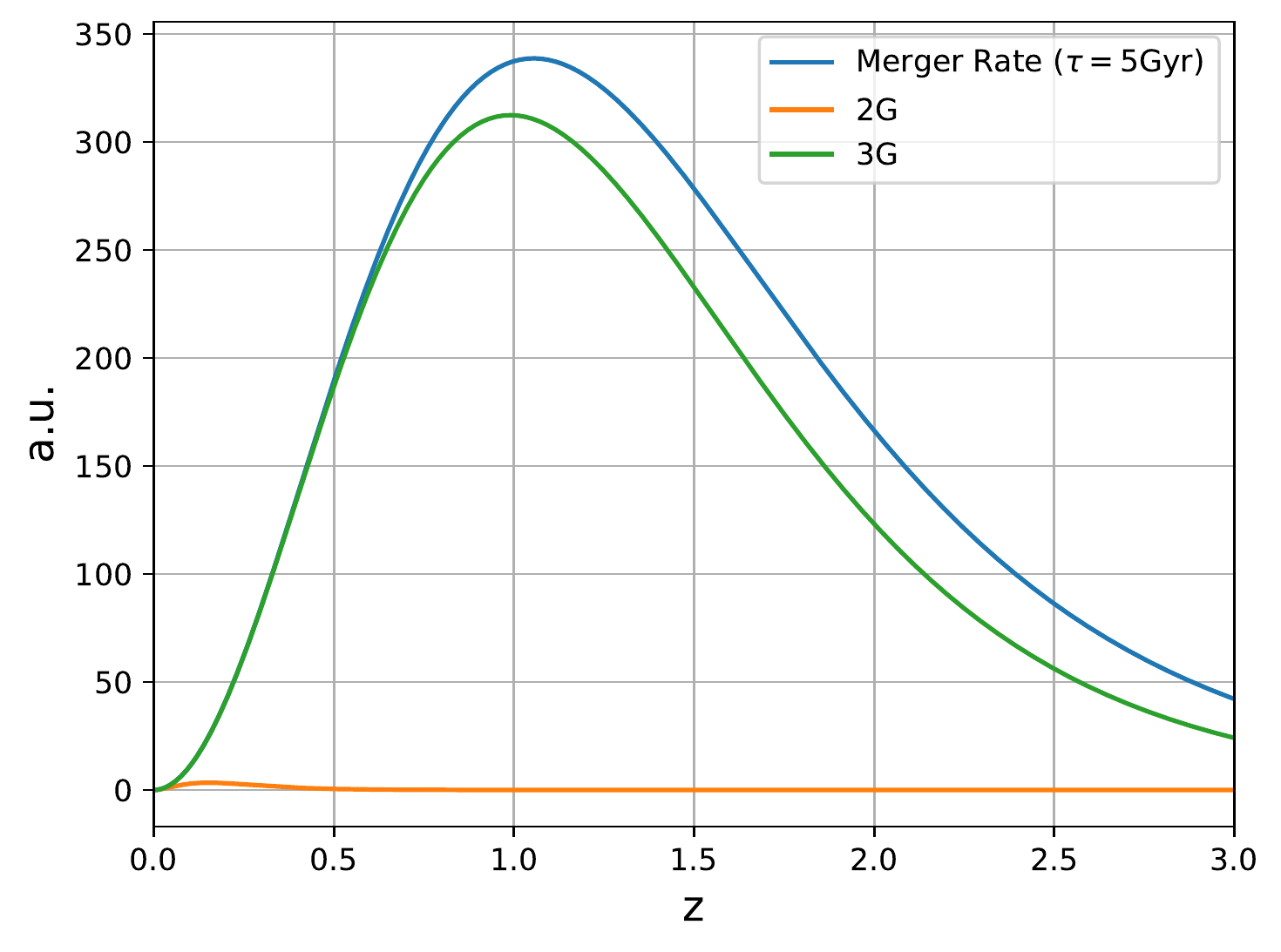}
  \end{center}
  \caption{Distribution of detections and merger rate (assuming $\tau=5$ Gyr).
    The 2G curve is normalized to unity, the 3G curve and merger rate have
    normalization consistent with the 2G curve.}
  \label{fig:dist}
\end{figure}

\begin{figure}
  \begin{center}
    \includegraphics[width=.9\columnwidth]{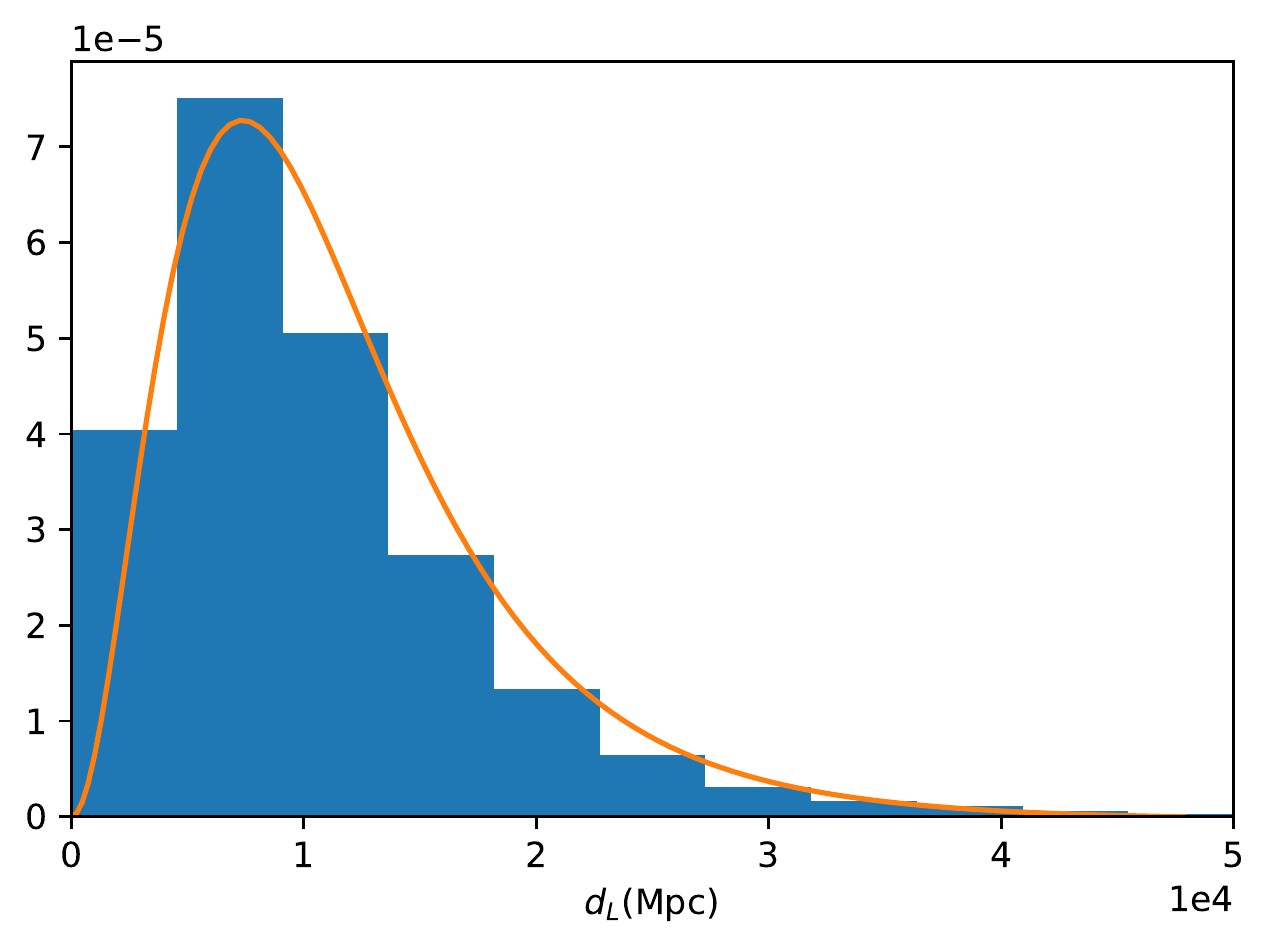}
  \end{center}
  \caption{Normalized distribution of simulated detections for a 3G
    detector.}
  \label{fig:injs_10000}
\end{figure}

We now perform the full analysis of Eq.~\eqref{postn}.
The merger rate of Eq.~(\ref{eq:merger_rate}) is represented in Fig.~\ref{fig:dist} for the fiducial values of $\tau=5$ Gyr, $H_0=69.32 \text{ km s}^{-1} {\rm Mpc}^{-1} $ and $\Omega_m=\Om^{(p)}$,
and for the detector sensitivities of 2G and 3G detectors (see Sec.~\ref{detector}).
We will now consider the case of the future 3G detectors. Fig.~\ref{fig:injs_10000} shows the normalized distribution of simulated injections for a 3G detector.

The expected absolute number of binary black hole observations by 3G detectors is poorly constrained because the underlying source distribution is known
only to a small extent.
By considering very different values of $\tau$ and normalizing the local merger rate density at $50$ Gpc$^{-3}$ yr$^{-1}$, one can see that, for instance, 10,000 detections can be accumulated in a time varying between a week
and few months  \cite{Vitale:2018yhm}.
Here, we consider the following possible scenarios -- 10,000, 20,000 and 40,000 detections -- which are realistic given the programmed duration of  future 3G observation runs.

\begin{figure}
  \begin{center}
    \includegraphics[width=\columnwidth]{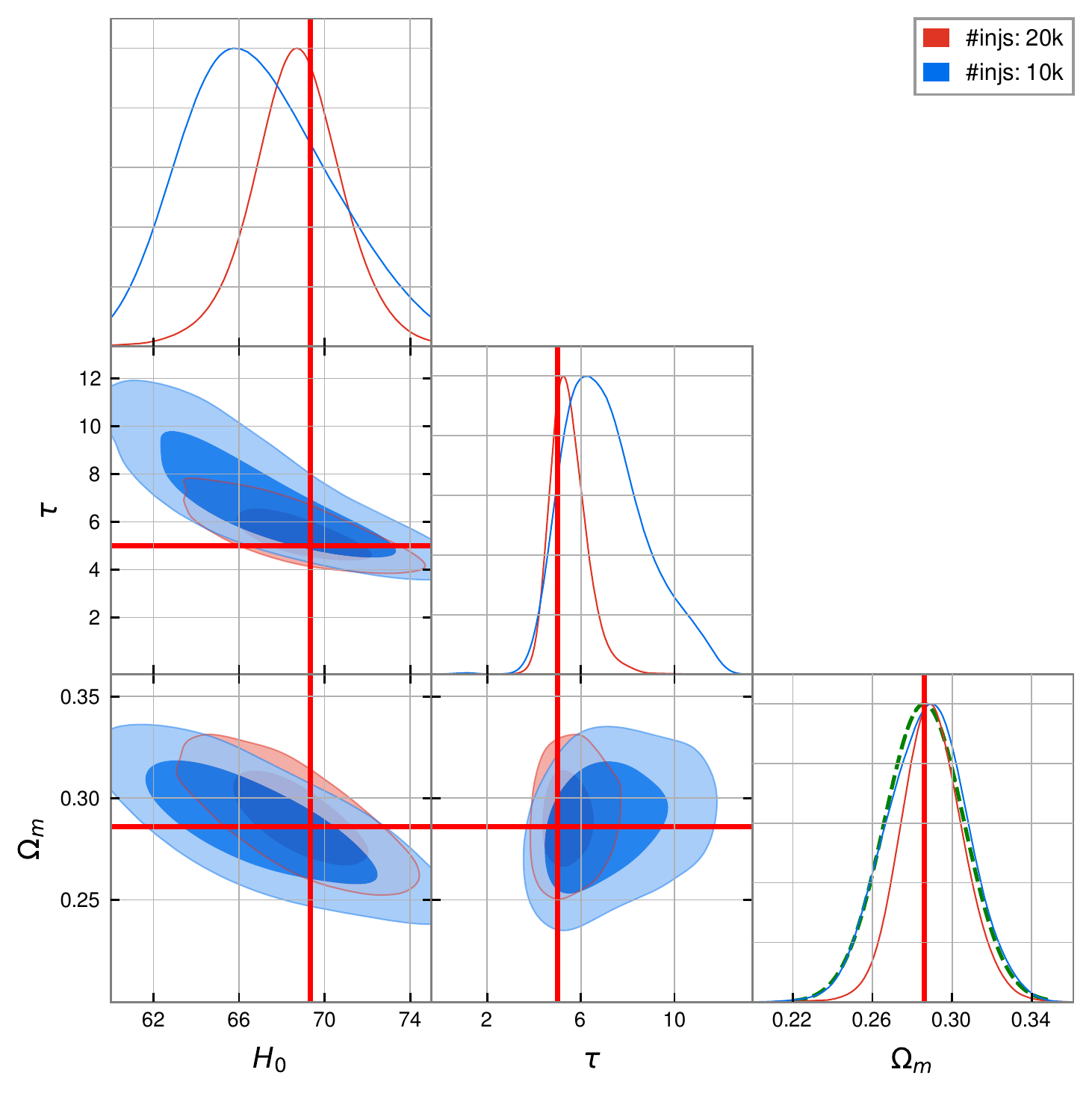}
  \end{center}
  \caption{Marginalized constraints on $H_0$, $\tau$ and $\Omega_m$ for 10,000
    and 20,000 simulated injections for the 3G case with 5\% relative errors in the measurement of $d_L$.
    The fiducial values of the parameters are marked with red lines.
    The prior on $\Omega_m$ is displayed with a green dashed line.}
  \label{fig:results3G}
\end{figure}

For the scenario with 10,000 and 20,000 injections, we fully sample the posterior via MCMC using the numerical codes \texttt{EMCEE} \cite{ForemanMackey:2012ig}, through its \texttt{Bilby} implementation
\cite{Ashton:2018jfp}, and \texttt{GETDIST} \cite{Lewis:2019xzd}.
The results are shown in Fig.~\ref{fig:results3G}, for the case of a 5\%
uncertainty in $d_L$, i.e. $\sigma_L/d_L=0.05$.
We can see that already with 10,000 GW observations it is possible to constrain the Hubble parameter at the few \% level.\\
As can be seen, the maximum of the posterior does not coincide exactly with the
fiducial value of the parameters (red lines in Fig.~\ref{fig:results3G}).
This is expected because in the present analysis it is not possible to perform
a forecast without fluctuations in the observational quantities. Indeed, while
one could fix the luminosity distances at their fiducial values,
the distribution in redshift of the injections is necessarily stochastic.
In other words, here we are considering fully realistic mock datasets.

\begin{figure}
  \begin{center}
    \includegraphics[width=\columnwidth]{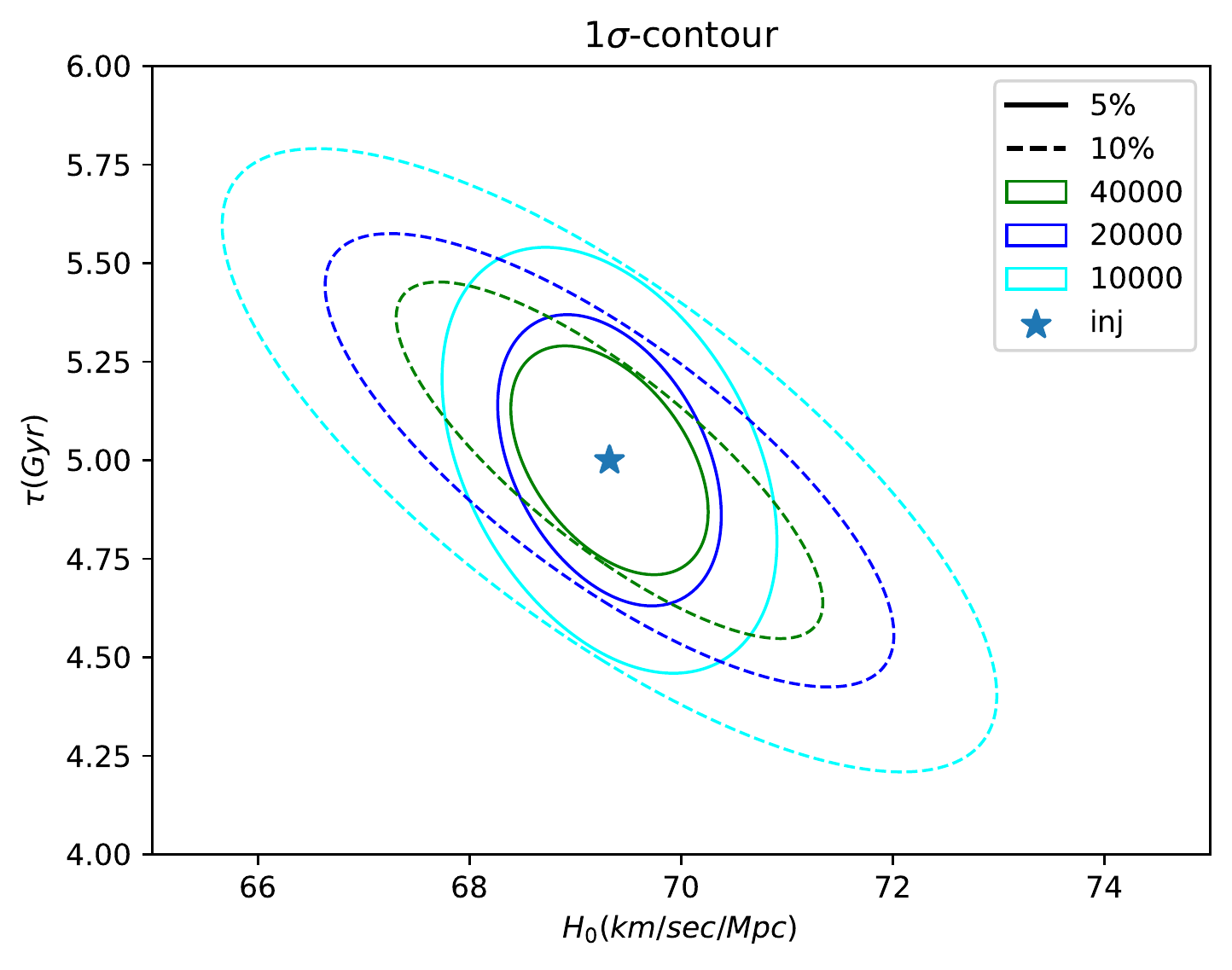}
  \end{center}
  \caption{1$\sigma$ covariance regions from averaged Fisher matrix.}
  \label{fig:covs_coh}
\end{figure}

Then, we analyze the scenario with 40,000 detections via the Fisher matrix approximation, obtained numerically via the \texttt{NUMDIFFTOOLS} library.%
\footnote{\href{https://pypi.org/project/numdifftools/}{pypi.org/project/numdifftools}.}
This is necessary because of the increased computational cost: as shown by Eq.~\eqref{postn} one has $n$ numerical integrals for $n$ injections. 
As explained earlier, the maximum of the posterior randomly walks around the fiducial value of the parameters and, to obtain a more robust estimate of the Fisher matrix against nonlinearities, we consider several sets of injections and average the corresponding Fisher matrices.
The result of this procedure is shown in Fig.~\ref{fig:covs_coh} (including also the cases that were analyzed via MCMC) and summarized in Table~\ref{calib} for the precision and Table~\ref{bias}
for the average bias in the recovered~$H_0$.
The results reported in Fig.~\ref{fig:results3G} give 1-$\sigma$ levels for
  $H_0$ of 5.5\% and 3.4\%  for 10,000 and 20,000 injections respectively, in agreement with the Fisher matrix estimations.

\begin{table}
  \begin{center}
    \setlength{\tabcolsep}{5pt}
    \renewcommand{\arraystretch}{1.5}
    \begin{tabular}{l>{\centering\arraybackslash}p{2.5cm}>{\centering\arraybackslash}p{2.5cm}}
      \hline
      \hline
      \multirow{2}{*}{\# injs} & \multicolumn{2}{c}{$\sigma_{H_0}/H_0$}\\
      &$\paq{\sigma_{d_L}/d_L=5\%}$ & $\paq{\sigma_{d_L}/d_L=10\%}$\\
      \hline
      10,000 &  4.9\% & 12.1\% \\
      20,000 &  3.0\% &  7.6\%  \\
      40,000 &  2.7\% &  6.5\%  \\
      \hline
      \hline
    \end{tabular}
    \caption{Forecasted relative constraints on $H_0$ for
      third-generation gravitational-wave detectors.}
    \label{calib}
  \end{center}
\end{table}

\begin{table}
  \begin{center}
    \setlength{\tabcolsep}{5pt}
    \renewcommand{\arraystretch}{1.5}
    \begin{tabular}{l>{\centering\arraybackslash}p{2.5cm}>{\centering\arraybackslash}p{2.5cm}}
      \hline
      \hline
      \multirow{2}{*}{\# injs} & \multicolumn{2}{c}{$\paq{\langle\pa{H_{0,{\rm inj}}/H_{0, {\rm rec}}-1}^2\rangle}^{1/2}$}\\
      &$\paq{\sigma_{d_L}/d_L=5\%}$ & $\paq{\sigma_{d_L}/d_L=10\%}$\\
      \hline
      10,000 & 2.7\% &  3.3\%  \\
      20,000 & 1.0\% &  1.0\%  \\
      40,000 & 0.5\% &  0.9\%  \\
      \hline
      \hline
    \end{tabular}
    \caption{Forecasted bias on $H_0$ for third-generation gravitational-wave
      detectors.}
    \label{bias}
  \end{center}
\end{table}

In the previous analysis we  assumed that one hyperparameter is enough to model our ignorance on the source distribution.
In the Appendix we show that one more hyperparameter can capture a possible bias in the adopted  start formation rate model.
This prevents the introduction of a bias in the Hubble constant at the price of degrading the precision of its determination, which worsens by a factor $\approx 2$.

\section{Conclusions} \label{conclusions}

Detections of gravitational waves from binary coalescences have opened
new ways to investigate cosmology.
In particular, while using concurrent observations of redshift and luminosity
distance is an obvious way to measure the Hubble constant, data from the first
three observation runs of LIGO and Virgo showed that binary black holes,
dark sirens without an electromagnetic counterpart, are far more frequent than
neutron star binaries with electromagnetic counterparts. 
Note, however, that forecasts for third generation detectors 
indicate that one could constrain the Hubble constant to subpercent level by accumulating electromagnetically bright standard sirens over
10 years at a rate of $\sim 30$ bright standard sirens per year \cite{Belgacem:2019tbw}.

On the other hand, by exploiting the gravitationally measured source location,
in the case of a network of at least three detectors, it has been shown that
already with $O(200)$ dark siren events one can achieve a few percent measurement of $H_0$ if the galaxy catalogs are at least 25\% complete \cite{Gray:2019ksv}.
This can be assumed only for relatively close sources, although galaxy catalogs complete
to magnitude 24 are expected to be produced by Euclid \cite{Blanchard:2019oqi},
allowing to see a Milky Way-type galaxy up to $1$ Gpc.

Here, we proposed an independent method, where
redshift information comes from our partial knowledge of the source
distribution. Marginalizing over the hyperparameter encoding our ignorance of
the binary astrophysical distributions we can estimate the Hubble
constant with a few percent precision with few tens of thousands 
black siren detections, without the need of multiple detectors, galaxy catalogs or
electromagnetic counterparts to have information about the individual source redshifts.
Note that, while the forecasted rate of binary black hole coalescence detections by third
generation gravitational wave observatories is subject to large
uncertainties, even in the more pessimistic scenarios few $O(10^3)$
detections per month should be made so that our method should be a viable alternative.

There are, however, caveats in our method.
First, to take into account detector-related selection effects, we have simulated future detections with a specific black hole mass
function. This will be addressed by the time our method will be used. Indeed, 3G detectors will have accumulated tens of thousands of BBH
detections so that we expect such mass function to be known accurately.
Second, the star formation rate we assumed may not correspond to the one
realized in nature and the model we presented in the main text, with only one
hyperparameter, may be an oversimplification. To test
these assumptions we have performed simulations in which data were injected and
analyzed using different star formation rate models.
The results reported in the Appendix show that the addition
of an another hyperparameter can capture the difference in underlying star
formation rate models and prevent the introduction of
a bias in the Hubble constant, though degrading the precision of its determination.

\section*{Acknowledgements}
The work of H.L. is financed in part by the Coordena\c{c}\~ao de Aperfei\c{c}oamento de Pessoal de N\'\i vel Superior - Brasil (CAPES) -- Finance Code 001.
V.M. thanks CNPq and FAPES for partial financial support.
R.S. thanks CNPq for partial financial support under Grant No.~312320/2018-3.
This project has received funding from the European Union’s Horizon 2020 research and innovation programme under the Marie Skłodowska-Curie Grant Agreement No. 888258.
We thank the High Performance Computing Center (NPAD) at UFRN for providing
computational resources.


%

\appendix

\section{Robustness against unknown star formation rate}
\label{ap:robu}

\begin{figure}
\begin{center}
\includegraphics[width=\columnwidth]{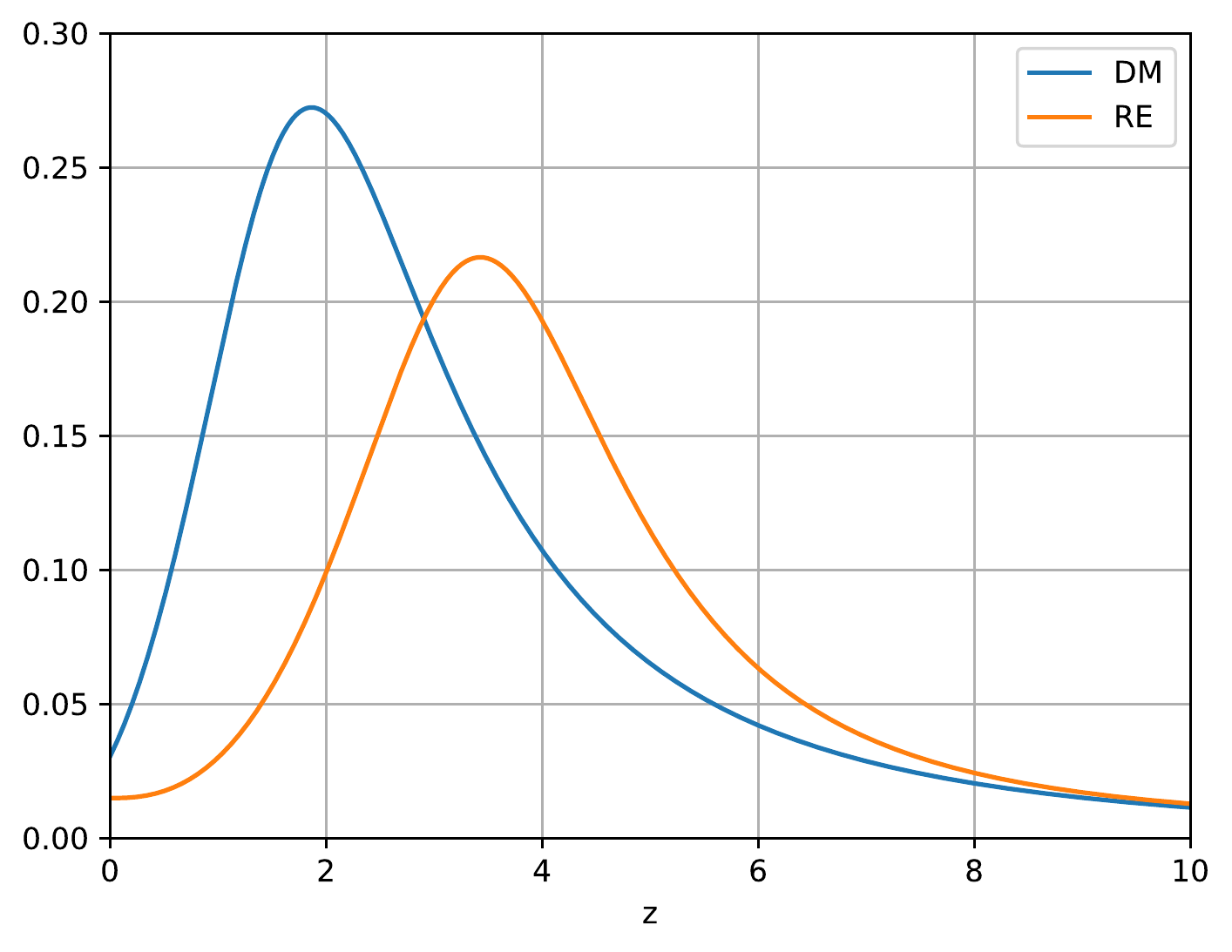}
\end{center}
\caption{The star formation rate of eqs.~(\ref{eq:sfrDM}) and (\ref{eq:sfrRE}).}
\label{fig:sfr2}
\end{figure}

\begin{figure}
\begin{center}
\includegraphics[width=\columnwidth]{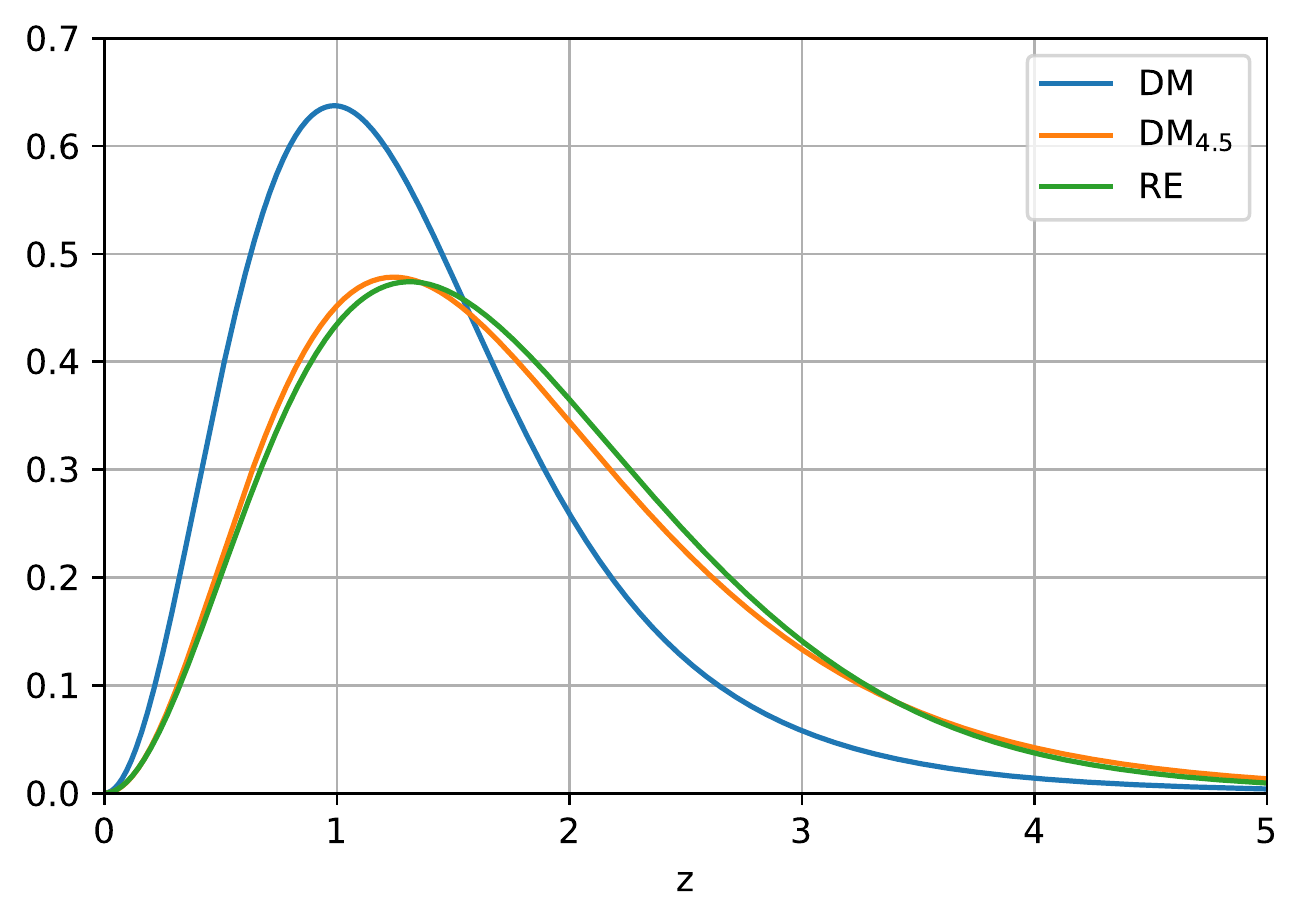}
\end{center}
\caption{Comparison of normalized detected merger rate assuming the DM star formation rate (\ref{eq:sfrDM}), the RE one (\ref{eq:sfrRE}), or the one of eq.~(\ref{eq:sfrDM}) with $C=4.5$.}
\label{fig:mr2}
\end{figure}

Our analysis adopts the star formation rate density $\psi_{\rm MD14}$ of equation~\eqref{eq:sfrDM} from \citet{Madau:2014bja}. Here, we investigate the impact of analyzing  with $\psi_{\rm MD14}$ data that were produced with the alternative star formation rate density by \citet{2012ApJ...744...95R}:
\begin{equation}
\label{eq:sfrRE}
  \psi_{\rm RE12}(z) \!=\! \!   \left [ \frac{0.007\!+\!0.27 (z/3.7)^{2.5}}{1\!+\!(z/3.7)^{6.4}} \!+\!0.003
   \right ]\! M_\odot \text{ yr}^{-1} \text{Mpc}^{-3}\,,
\end{equation}
to have a proxy of the bias we may introduce in the cosmological parameter estimation by adopting an incorrect underlying star formation and merger distribution. Both functions are plotted in Fig.~\ref{fig:sfr2}.
%
Fig.~\ref{fig:mr2} shows that despite the two underlying star formation rates
are qualitatively different, the resulting merger rates can be made to overlap
by adjusting the $C$ parameter of Eq.~(\ref{eq:sfrDM}), which we now promote to hyper-parameter (and treat as a nuisance parameter).

\begin{figure}
\begin{center}
  \includegraphics[width=\columnwidth]{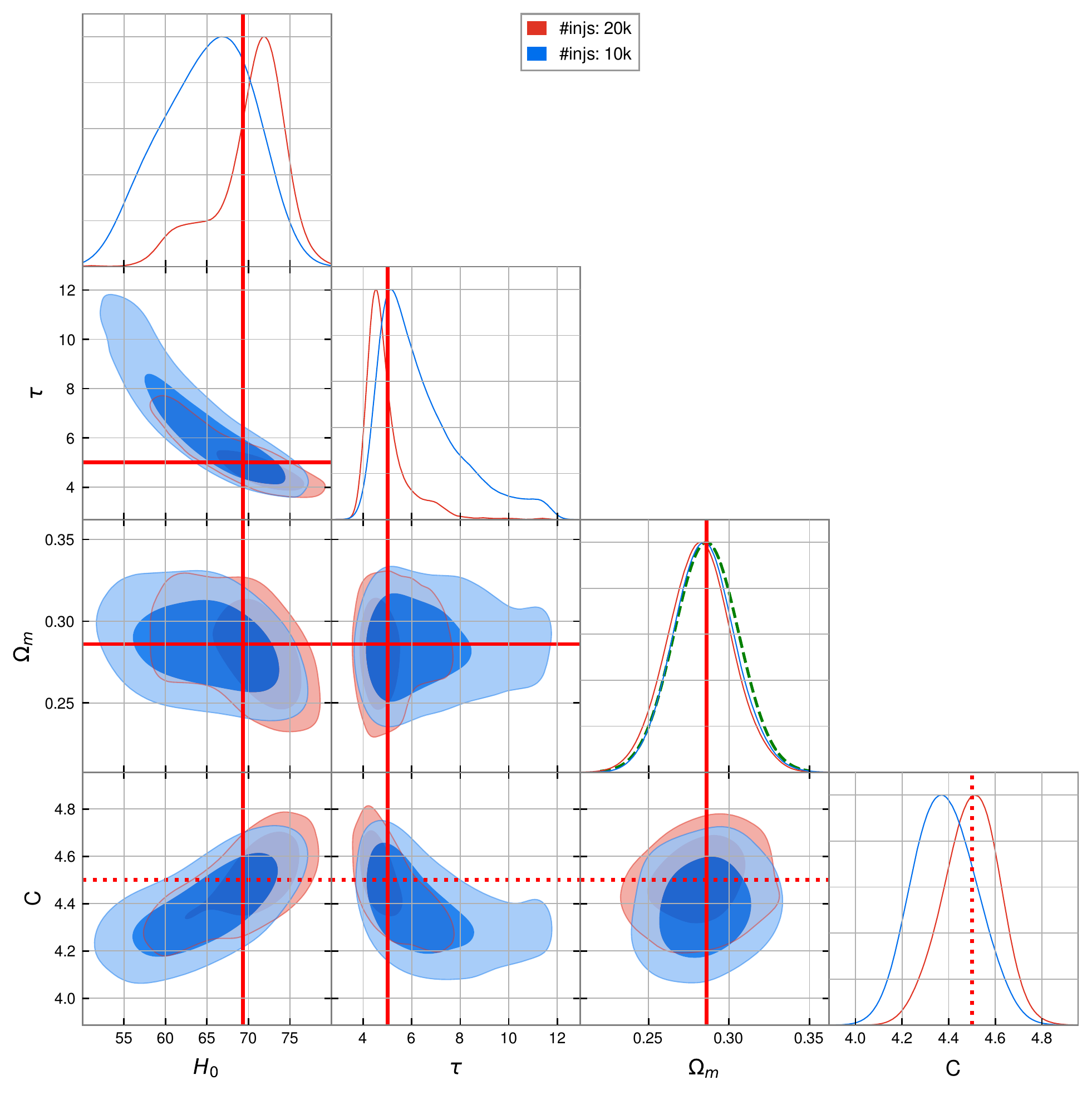}
\end{center}
\caption{Statistical inference for injections generated according to the star
  formation rate of Eq.~(\ref{eq:sfrRE}) but analyzed using the star formation
  rate of Eq.~(\ref{eq:sfrDM}), for 5\% relative errors in the measurement of $d_L$, with 20,000 and 10,000 injections.
  Here, the model includes the nuisance parameter $C$, which absorbs the effect
  of a different star formation rate between injection and recovery,
  with the result of keeping $H_0$ unbiased.}
\label{fig:4pars}
\end{figure}

We then show in Fig.~\ref{fig:4pars} the results of an analysis in which
the probability distributions for $H_0,\ \Omega_m$ and the two nuisance parameters
$\tau$ and $C$ are obtained in the case in which the injections are generated assuming the star formation rate (\ref{eq:sfrRE}) but analyzed with the star formation rate (\ref{eq:sfrDM}).
One can see that the hyper-parameter $C$, by taking a value different from the original one of eq.~(\ref{eq:sfrDM}), absorbs the effect of a different star formation rate, avoiding a bias in $H_0$.
On the other hand, the precision on $H_0$ is degraded to almost 10\% percent level, thus
requiring several tens of thousand of injections to reach percent level.

\begin{thebibliography}{58}%
\makeatletter
\providecommand \@ifxundefined [1]{%
 \@ifx{#1\undefined}
}%
\providecommand \@ifnum [1]{%
 \ifnum #1\expandafter \@firstoftwo
 \else \expandafter \@secondoftwo
 \fi
}%
\providecommand \@ifx [1]{%
 \ifx #1\expandafter \@firstoftwo
 \else \expandafter \@secondoftwo
 \fi
}%
\providecommand \natexlab [1]{#1}%
\providecommand \enquote  [1]{``#1''}%
\providecommand \bibnamefont  [1]{#1}%
\providecommand \bibfnamefont [1]{#1}%
\providecommand \citenamefont [1]{#1}%
\providecommand \href@noop [0]{\@secondoftwo}%
\providecommand \href [0]{\begingroup \@sanitize@url \@href}%
\providecommand \@href[1]{\@@startlink{#1}\@@href}%
\providecommand \@@href[1]{\endgroup#1\@@endlink}%
\providecommand \@sanitize@url [0]{\catcode `\\12\catcode `\$12\catcode
  `\&12\catcode `\#12\catcode `\^12\catcode `\_12\catcode `\%12\relax}%
\providecommand \@@startlink[1]{}%
\providecommand \@@endlink[0]{}%
\providecommand \url  [0]{\begingroup\@sanitize@url \@url }%
\providecommand \@url [1]{\endgroup\@href {#1}{\urlprefix }}%
\providecommand \urlprefix  [0]{URL }%
\providecommand \Eprint [0]{\href }%
\providecommand \doibase [0]{http://dx.doi.org/}%
\providecommand \selectlanguage [0]{\@gobble}%
\providecommand \bibinfo  [0]{\@secondoftwo}%
\providecommand \bibfield  [0]{\@secondoftwo}%
\providecommand \translation [1]{[#1]}%
\providecommand \BibitemOpen [0]{}%
\providecommand \bibitemStop [0]{}%
\providecommand \bibitemNoStop [0]{.\EOS\space}%
\providecommand \EOS [0]{\spacefactor3000\relax}%
\providecommand \BibitemShut  [1]{\csname bibitem#1\endcsname}%
\let\auto@bib@innerbib\@empty
\bibitem [{\citenamefont {Riess}\ \emph {et~al.}(2021)\citenamefont {Riess},
  \citenamefont {Casertano}, \citenamefont {Yuan}, \citenamefont {Bowers},
  \citenamefont {Macri}, \citenamefont {Zinn},\ and\ \citenamefont
  {Scolnic}}]{Riess:2020fzl}%
  \BibitemOpen
  \bibfield  {author} {\bibinfo {author} {\bibfnamefont {A.~G.}\ \bibnamefont
  {Riess}}, \bibinfo {author} {\bibfnamefont {S.}~\bibnamefont {Casertano}},
  \bibinfo {author} {\bibfnamefont {W.}~\bibnamefont {Yuan}}, \bibinfo {author}
  {\bibfnamefont {J.~B.}\ \bibnamefont {Bowers}}, \bibinfo {author}
  {\bibfnamefont {L.}~\bibnamefont {Macri}}, \bibinfo {author} {\bibfnamefont
  {J.~C.}\ \bibnamefont {Zinn}}, \ and\ \bibinfo {author} {\bibfnamefont
  {D.}~\bibnamefont {Scolnic}},\ }\href {\doibase 10.3847/2041-8213/abdbaf}
  {\bibfield  {journal} {\bibinfo  {journal} {Astrophys. J. Lett.}\ }\textbf
  {\bibinfo {volume} {908}},\ \bibinfo {pages} {L6} (\bibinfo {year} {2021})},\
  \Eprint {http://arxiv.org/abs/2012.08534} {arXiv:2012.08534 [astro-ph.CO]}
  \BibitemShut {NoStop}%
\bibitem [{\citenamefont {Aghanim}\ \emph {et~al.}(2020)\citenamefont {Aghanim}
  \emph {et~al.}}]{Aghanim:2018eyx}%
  \BibitemOpen
  \bibfield  {author} {\bibinfo {author} {\bibfnamefont {N.}~\bibnamefont
  {Aghanim}} \emph {et~al.} (\bibinfo {collaboration} {Planck}),\ }\href
  {\doibase 10.1051/0004-6361/201833910} {\bibfield  {journal} {\bibinfo
  {journal} {Astron. Astrophys.}\ }\textbf {\bibinfo {volume} {641}},\ \bibinfo
  {pages} {A6} (\bibinfo {year} {2020})},\ \Eprint
  {http://arxiv.org/abs/1807.06209} {arXiv:1807.06209 [astro-ph.CO]}
  \BibitemShut {NoStop}%
\bibitem [{\citenamefont {Camarena}\ and\ \citenamefont
  {Marra}(2021)}]{Camarena:2021jlr}%
  \BibitemOpen
  \bibfield  {author} {\bibinfo {author} {\bibfnamefont {D.}~\bibnamefont
  {Camarena}}\ and\ \bibinfo {author} {\bibfnamefont {V.}~\bibnamefont
  {Marra}},\ }\href {\doibase 10.1093/mnras/stab1200} {\bibfield  {journal}
  {\bibinfo  {journal} {Mon. Not. Roy. Astron. Soc.}\ }\textbf {\bibinfo
  {volume} {{in press}}} (\bibinfo {year} {2021}),\ 10.1093/mnras/stab1200},\
  \Eprint {http://arxiv.org/abs/2101.08641} {arXiv:2101.08641 [astro-ph.CO]}
  \BibitemShut {NoStop}%
\bibitem [{\citenamefont {Knox}\ and\ \citenamefont
  {Millea}(2020)}]{Knox:2019rjx}%
  \BibitemOpen
  \bibfield  {author} {\bibinfo {author} {\bibfnamefont {L.}~\bibnamefont
  {Knox}}\ and\ \bibinfo {author} {\bibfnamefont {M.}~\bibnamefont {Millea}},\
  }\href {\doibase 10.1103/PhysRevD.101.043533} {\bibfield  {journal} {\bibinfo
   {journal} {Phys. Rev. D}\ }\textbf {\bibinfo {volume} {101}},\ \bibinfo
  {pages} {043533} (\bibinfo {year} {2020})},\ \Eprint
  {http://arxiv.org/abs/1908.03663} {arXiv:1908.03663 [astro-ph.CO]}
  \BibitemShut {NoStop}%
\bibitem [{\citenamefont {Verde}\ \emph {et~al.}(2019)\citenamefont {Verde},
  \citenamefont {Treu},\ and\ \citenamefont {Riess}}]{Verde:2019ivm}%
  \BibitemOpen
  \bibfield  {author} {\bibinfo {author} {\bibfnamefont {L.}~\bibnamefont
  {Verde}}, \bibinfo {author} {\bibfnamefont {T.}~\bibnamefont {Treu}}, \ and\
  \bibinfo {author} {\bibfnamefont {A.}~\bibnamefont {Riess}}\ }(\bibinfo
  {year} {2019})\ \Eprint {http://arxiv.org/abs/1907.10625} {arXiv:1907.10625
  [astro-ph.CO]} \BibitemShut {NoStop}%
\bibitem [{\citenamefont {Di~Valentino}\ \emph {et~al.}(2021)\citenamefont
  {Di~Valentino}, \citenamefont {Mena}, \citenamefont {Pan}, \citenamefont
  {Visinelli}, \citenamefont {Yang}, \citenamefont {Melchiorri}, \citenamefont
  {Mota}, \citenamefont {Riess},\ and\ \citenamefont
  {Silk}}]{DiValentino:2021izs}%
  \BibitemOpen
  \bibfield  {author} {\bibinfo {author} {\bibfnamefont {E.}~\bibnamefont
  {Di~Valentino}}, \bibinfo {author} {\bibfnamefont {O.}~\bibnamefont {Mena}},
  \bibinfo {author} {\bibfnamefont {S.}~\bibnamefont {Pan}}, \bibinfo {author}
  {\bibfnamefont {L.}~\bibnamefont {Visinelli}}, \bibinfo {author}
  {\bibfnamefont {W.}~\bibnamefont {Yang}}, \bibinfo {author} {\bibfnamefont
  {A.}~\bibnamefont {Melchiorri}}, \bibinfo {author} {\bibfnamefont {D.~F.}\
  \bibnamefont {Mota}}, \bibinfo {author} {\bibfnamefont {A.~G.}\ \bibnamefont
  {Riess}}, \ and\ \bibinfo {author} {\bibfnamefont {J.}~\bibnamefont {Silk}},\
  }\href {\doibase 10.1088/1361-6382/ac086d} {\bibfield  {journal} {\bibinfo
  {journal} {Class. Quant. Grav.}\ }\textbf {\bibinfo {volume} {38}},\ \bibinfo
  {pages} {153001} (\bibinfo {year} {2021})},\ \Eprint
  {http://arxiv.org/abs/2103.01183} {arXiv:2103.01183 [astro-ph.CO]}
  \BibitemShut {NoStop}%
\bibitem [{\citenamefont {Perivolaropoulos}\ and\ \citenamefont
  {Skara}(2021)}]{Perivolaropoulos:2021jda}%
  \BibitemOpen
  \bibfield  {author} {\bibinfo {author} {\bibfnamefont {L.}~\bibnamefont
  {Perivolaropoulos}}\ and\ \bibinfo {author} {\bibfnamefont {F.}~\bibnamefont
  {Skara}},\ }\href@noop {} {\  (\bibinfo {year} {2021})},\ \Eprint
  {http://arxiv.org/abs/2105.05208} {arXiv:2105.05208 [astro-ph.CO]}
  \BibitemShut {NoStop}%
\bibitem [{\citenamefont {Khetan}\ \emph {et~al.}(2021)\citenamefont {Khetan}
  \emph {et~al.}}]{Khetan:2020hmh}%
  \BibitemOpen
  \bibfield  {author} {\bibinfo {author} {\bibfnamefont {N.}~\bibnamefont
  {Khetan}} \emph {et~al.},\ }\href {\doibase 10.1051/0004-6361/202039196}
  {\bibfield  {journal} {\bibinfo  {journal} {Astron. Astrophys.}\ }\textbf
  {\bibinfo {volume} {647}},\ \bibinfo {pages} {A72} (\bibinfo {year}
  {2021})},\ \Eprint {http://arxiv.org/abs/2008.07754} {arXiv:2008.07754
  [astro-ph.CO]} \BibitemShut {NoStop}%
\bibitem [{\citenamefont {Gray}\ \emph {et~al.}(2020)\citenamefont {Gray} \emph
  {et~al.}}]{Gray:2019ksv}%
  \BibitemOpen
  \bibfield  {author} {\bibinfo {author} {\bibfnamefont {R.}~\bibnamefont
  {Gray}} \emph {et~al.},\ }\href {\doibase 10.1103/PhysRevD.101.122001}
  {\bibfield  {journal} {\bibinfo  {journal} {Phys. Rev. D}\ }\textbf {\bibinfo
  {volume} {101}},\ \bibinfo {pages} {122001} (\bibinfo {year} {2020})},\
  \Eprint {http://arxiv.org/abs/1908.06050} {arXiv:1908.06050 [gr-qc]}
  \BibitemShut {NoStop}%
\bibitem [{\citenamefont {Aasi}\ \emph {et~al.}(2015)\citenamefont {Aasi} \emph
  {et~al.}}]{TheLIGOScientific:2014jea}%
  \BibitemOpen
  \bibfield  {author} {\bibinfo {author} {\bibfnamefont {J.}~\bibnamefont
  {Aasi}} \emph {et~al.} (\bibinfo {collaboration} {LIGO Scientific}),\ }\href
  {\doibase 10.1088/0264-9381/32/7/074001} {\bibfield  {journal} {\bibinfo
  {journal} {Class. Quant. Grav.}\ }\textbf {\bibinfo {volume} {32}},\ \bibinfo
  {pages} {074001} (\bibinfo {year} {2015})},\ \Eprint
  {http://arxiv.org/abs/1411.4547} {arXiv:1411.4547 [gr-qc]} \BibitemShut
  {NoStop}%
\bibitem [{\citenamefont {Acernese}\ \emph {et~al.}(2015)\citenamefont
  {Acernese} \emph {et~al.}}]{TheVirgo:2014hva}%
  \BibitemOpen
  \bibfield  {author} {\bibinfo {author} {\bibfnamefont {F.}~\bibnamefont
  {Acernese}} \emph {et~al.} (\bibinfo {collaboration} {VIRGO}),\ }\href
  {\doibase 10.1088/0264-9381/32/2/024001} {\bibfield  {journal} {\bibinfo
  {journal} {Class. Quant. Grav.}\ }\textbf {\bibinfo {volume} {32}},\ \bibinfo
  {pages} {024001} (\bibinfo {year} {2015})},\ \Eprint
  {http://arxiv.org/abs/1408.3978} {arXiv:1408.3978 [gr-qc]} \BibitemShut
  {NoStop}%
\bibitem [{\citenamefont {Akutsu}\ \emph {et~al.}(2021)\citenamefont {Akutsu}
  \emph {et~al.}}]{KAGRA:2020tym}%
  \BibitemOpen
  \bibfield  {author} {\bibinfo {author} {\bibfnamefont {T.}~\bibnamefont
  {Akutsu}} \emph {et~al.} (\bibinfo {collaboration} {KAGRA}),\ }\href
  {\doibase 10.1093/ptep/ptaa125} {\bibfield  {journal} {\bibinfo  {journal}
  {PTEP}\ }\textbf {\bibinfo {volume} {2021}},\ \bibinfo {pages} {05A101}
  (\bibinfo {year} {2021})},\ \Eprint {http://arxiv.org/abs/2005.05574}
  {arXiv:2005.05574 [physics.ins-det]} \BibitemShut {NoStop}%
\bibitem [{\citenamefont {Punturo}\ \emph {et~al.}(2010)\citenamefont {Punturo}
  \emph {et~al.}}]{Punturo:2010zz}%
  \BibitemOpen
  \bibfield  {author} {\bibinfo {author} {\bibfnamefont {M.}~\bibnamefont
  {Punturo}} \emph {et~al.},\ }\href {\doibase 10.1088/0264-9381/27/19/194002}
  {\bibfield  {journal} {\bibinfo  {journal} {Class. Quant. Grav.}\ }\textbf
  {\bibinfo {volume} {27}},\ \bibinfo {pages} {194002} (\bibinfo {year}
  {2010})}\BibitemShut {NoStop}%
\bibitem [{\citenamefont {Abbott}\ \emph
  {et~al.}(2017{\natexlab{a}})\citenamefont {Abbott} \emph
  {et~al.}}]{Evans:2016mbw}%
  \BibitemOpen
  \bibfield  {author} {\bibinfo {author} {\bibfnamefont {B.~P.}\ \bibnamefont
  {Abbott}} \emph {et~al.} (\bibinfo {collaboration} {LIGO Scientific}),\
  }\href {\doibase 10.1088/1361-6382/aa51f4} {\bibfield  {journal} {\bibinfo
  {journal} {Class. Quant. Grav.}\ }\textbf {\bibinfo {volume} {34}},\ \bibinfo
  {pages} {044001} (\bibinfo {year} {2017}{\natexlab{a}})},\ \Eprint
  {http://arxiv.org/abs/1607.08697} {arXiv:1607.08697 [astro-ph.IM]}
  \BibitemShut {NoStop}%
\bibitem [{\citenamefont {Abbott}\ \emph {et~al.}(2020)\citenamefont {Abbott}
  \emph {et~al.}}]{KAGRA:2020npa}%
  \BibitemOpen
  \bibfield  {author} {\bibinfo {author} {\bibfnamefont {B.~P.}\ \bibnamefont
  {Abbott}} \emph {et~al.} (\bibinfo {collaboration} {KAGRA, LIGO Scientific,
  Virgo}),\ }\href {\doibase 10.1007/s41114-020-00026-9} {\bibfield  {journal}
  {\bibinfo  {journal} {Living Rev. Rel.}\ }\textbf {\bibinfo {volume} {23}},\
  \bibinfo {pages} {3} (\bibinfo {year} {2020})}\BibitemShut {NoStop}%
\bibitem [{\citenamefont {Camarena}\ and\ \citenamefont
  {Marra}(2018)}]{Camarena:2018nbr}%
  \BibitemOpen
  \bibfield  {author} {\bibinfo {author} {\bibfnamefont {D.}~\bibnamefont
  {Camarena}}\ and\ \bibinfo {author} {\bibfnamefont {V.}~\bibnamefont
  {Marra}},\ }\href {\doibase 10.1103/PhysRevD.98.023537} {\bibfield  {journal}
  {\bibinfo  {journal} {Phys. Rev.}\ }\textbf {\bibinfo {volume} {D98}},\
  \bibinfo {pages} {023537} (\bibinfo {year} {2018})},\ \Eprint
  {http://arxiv.org/abs/1805.09900} {arXiv:1805.09900 [astro-ph.CO]}
  \BibitemShut {NoStop}%
\bibitem [{\citenamefont {Schutz}(1986)}]{Schutz:1986gp}%
  \BibitemOpen
  \bibfield  {author} {\bibinfo {author} {\bibfnamefont {B.~F.}\ \bibnamefont
  {Schutz}},\ }\href {\doibase 10.1038/323310a0} {\bibfield  {journal}
  {\bibinfo  {journal} {Nature}\ }\textbf {\bibinfo {volume} {323}},\ \bibinfo
  {pages} {310} (\bibinfo {year} {1986})}\BibitemShut {NoStop}%
\bibitem [{\citenamefont {Holz}\ and\ \citenamefont
  {Hughes}(2005)}]{Holz:2005df}%
  \BibitemOpen
  \bibfield  {author} {\bibinfo {author} {\bibfnamefont {D.~E.}\ \bibnamefont
  {Holz}}\ and\ \bibinfo {author} {\bibfnamefont {S.~A.}\ \bibnamefont
  {Hughes}},\ }\href {\doibase 10.1086/431341} {\bibfield  {journal} {\bibinfo
  {journal} {Astrophys. J.}\ }\textbf {\bibinfo {volume} {629}},\ \bibinfo
  {pages} {15} (\bibinfo {year} {2005})},\ \Eprint
  {http://arxiv.org/abs/astro-ph/0504616} {arXiv:astro-ph/0504616} \BibitemShut
  {NoStop}%
\bibitem [{\citenamefont {Chen}\ \emph {et~al.}(2018)\citenamefont {Chen},
  \citenamefont {Fishbach},\ and\ \citenamefont {Holz}}]{Chen:2017rfc}%
  \BibitemOpen
  \bibfield  {author} {\bibinfo {author} {\bibfnamefont {H.-Y.}\ \bibnamefont
  {Chen}}, \bibinfo {author} {\bibfnamefont {M.}~\bibnamefont {Fishbach}}, \
  and\ \bibinfo {author} {\bibfnamefont {D.~E.}\ \bibnamefont {Holz}},\ }\href
  {\doibase 10.1038/s41586-018-0606-0} {\bibfield  {journal} {\bibinfo
  {journal} {Nature}\ }\textbf {\bibinfo {volume} {562}},\ \bibinfo {pages}
  {545} (\bibinfo {year} {2018})},\ \Eprint {http://arxiv.org/abs/1712.06531}
  {arXiv:1712.06531 [astro-ph.CO]} \BibitemShut {NoStop}%
\bibitem [{\citenamefont {Dalal}\ \emph {et~al.}(2006)\citenamefont {Dalal},
  \citenamefont {Holz}, \citenamefont {Hughes},\ and\ \citenamefont
  {Jain}}]{Dalal:2006qt}%
  \BibitemOpen
  \bibfield  {author} {\bibinfo {author} {\bibfnamefont {N.}~\bibnamefont
  {Dalal}}, \bibinfo {author} {\bibfnamefont {D.~E.}\ \bibnamefont {Holz}},
  \bibinfo {author} {\bibfnamefont {S.~A.}\ \bibnamefont {Hughes}}, \ and\
  \bibinfo {author} {\bibfnamefont {B.}~\bibnamefont {Jain}},\ }\href {\doibase
  10.1103/PhysRevD.74.063006} {\bibfield  {journal} {\bibinfo  {journal} {Phys.
  Rev. D}\ }\textbf {\bibinfo {volume} {74}},\ \bibinfo {pages} {063006}
  (\bibinfo {year} {2006})},\ \Eprint {http://arxiv.org/abs/astro-ph/0601275}
  {arXiv:astro-ph/0601275} \BibitemShut {NoStop}%
\bibitem [{\citenamefont {Nissanke}\ \emph {et~al.}(2010)\citenamefont
  {Nissanke}, \citenamefont {Holz}, \citenamefont {Hughes}, \citenamefont
  {Dalal},\ and\ \citenamefont {Sievers}}]{Nissanke:2009kt}%
  \BibitemOpen
  \bibfield  {author} {\bibinfo {author} {\bibfnamefont {S.}~\bibnamefont
  {Nissanke}}, \bibinfo {author} {\bibfnamefont {D.~E.}\ \bibnamefont {Holz}},
  \bibinfo {author} {\bibfnamefont {S.~A.}\ \bibnamefont {Hughes}}, \bibinfo
  {author} {\bibfnamefont {N.}~\bibnamefont {Dalal}}, \ and\ \bibinfo {author}
  {\bibfnamefont {J.~L.}\ \bibnamefont {Sievers}},\ }\href {\doibase
  10.1088/0004-637X/725/1/496} {\bibfield  {journal} {\bibinfo  {journal}
  {Astrophys. J.}\ }\textbf {\bibinfo {volume} {725}},\ \bibinfo {pages} {496}
  (\bibinfo {year} {2010})},\ \Eprint {http://arxiv.org/abs/0904.1017}
  {arXiv:0904.1017 [astro-ph.CO]} \BibitemShut {NoStop}%
\bibitem [{\citenamefont {Belgacem}\ \emph {et~al.}(2019)\citenamefont
  {Belgacem}, \citenamefont {Dirian}, \citenamefont {Foffa}, \citenamefont
  {Howell}, \citenamefont {Maggiore},\ and\ \citenamefont
  {Regimbau}}]{Belgacem:2019tbw}%
  \BibitemOpen
  \bibfield  {author} {\bibinfo {author} {\bibfnamefont {E.}~\bibnamefont
  {Belgacem}}, \bibinfo {author} {\bibfnamefont {Y.}~\bibnamefont {Dirian}},
  \bibinfo {author} {\bibfnamefont {S.}~\bibnamefont {Foffa}}, \bibinfo
  {author} {\bibfnamefont {E.~J.}\ \bibnamefont {Howell}}, \bibinfo {author}
  {\bibfnamefont {M.}~\bibnamefont {Maggiore}}, \ and\ \bibinfo {author}
  {\bibfnamefont {T.}~\bibnamefont {Regimbau}},\ }\href {\doibase
  10.1088/1475-7516/2019/08/015} {\bibfield  {journal} {\bibinfo  {journal}
  {JCAP}\ }\textbf {\bibinfo {volume} {08}},\ \bibinfo {pages} {015} (\bibinfo
  {year} {2019})},\ \Eprint {http://arxiv.org/abs/1907.01487} {arXiv:1907.01487
  [astro-ph.CO]} \BibitemShut {NoStop}%
\bibitem [{\citenamefont {Zhang}\ \emph {et~al.}(2019)\citenamefont {Zhang},
  \citenamefont {Zhang}, \citenamefont {Jin}, \citenamefont {Qi},\ and\
  \citenamefont {Zhang}}]{Zhang:2019loq}%
  \BibitemOpen
  \bibfield  {author} {\bibinfo {author} {\bibfnamefont {J.-F.}\ \bibnamefont
  {Zhang}}, \bibinfo {author} {\bibfnamefont {M.}~\bibnamefont {Zhang}},
  \bibinfo {author} {\bibfnamefont {S.-J.}\ \bibnamefont {Jin}}, \bibinfo
  {author} {\bibfnamefont {J.-Z.}\ \bibnamefont {Qi}}, \ and\ \bibinfo {author}
  {\bibfnamefont {X.}~\bibnamefont {Zhang}},\ }\href {\doibase
  10.1088/1475-7516/2019/09/068} {\bibfield  {journal} {\bibinfo  {journal}
  {JCAP}\ }\textbf {\bibinfo {volume} {09}},\ \bibinfo {pages} {068} (\bibinfo
  {year} {2019})},\ \Eprint {http://arxiv.org/abs/1907.03238} {arXiv:1907.03238
  [astro-ph.CO]} \BibitemShut {NoStop}%
\bibitem [{\citenamefont {Jin}\ \emph {et~al.}(2020)\citenamefont {Jin},
  \citenamefont {He}, \citenamefont {Xu}, \citenamefont {Zhang},\ and\
  \citenamefont {Zhang}}]{Jin:2020hmc}%
  \BibitemOpen
  \bibfield  {author} {\bibinfo {author} {\bibfnamefont {S.-J.}\ \bibnamefont
  {Jin}}, \bibinfo {author} {\bibfnamefont {D.-Z.}\ \bibnamefont {He}},
  \bibinfo {author} {\bibfnamefont {Y.}~\bibnamefont {Xu}}, \bibinfo {author}
  {\bibfnamefont {J.-F.}\ \bibnamefont {Zhang}}, \ and\ \bibinfo {author}
  {\bibfnamefont {X.}~\bibnamefont {Zhang}},\ }\href {\doibase
  10.1088/1475-7516/2020/03/051} {\bibfield  {journal} {\bibinfo  {journal}
  {JCAP}\ }\textbf {\bibinfo {volume} {03}},\ \bibinfo {pages} {051} (\bibinfo
  {year} {2020})},\ \Eprint {http://arxiv.org/abs/2001.05393} {arXiv:2001.05393
  [astro-ph.CO]} \BibitemShut {NoStop}%
\bibitem [{\citenamefont {Jin}\ \emph {et~al.}(2021)\citenamefont {Jin},
  \citenamefont {Wang}, \citenamefont {Wu}, \citenamefont {Zhang},\ and\
  \citenamefont {Zhang}}]{Jin:2021pcv}%
  \BibitemOpen
  \bibfield  {author} {\bibinfo {author} {\bibfnamefont {S.-J.}\ \bibnamefont
  {Jin}}, \bibinfo {author} {\bibfnamefont {L.-F.}\ \bibnamefont {Wang}},
  \bibinfo {author} {\bibfnamefont {P.-J.}\ \bibnamefont {Wu}}, \bibinfo
  {author} {\bibfnamefont {J.-F.}\ \bibnamefont {Zhang}}, \ and\ \bibinfo
  {author} {\bibfnamefont {X.}~\bibnamefont {Zhang}},\ }\href {\doibase
  10.1103/PhysRevD.104.103507} {\bibfield  {journal} {\bibinfo  {journal}
  {Phys. Rev. D}\ }\textbf {\bibinfo {volume} {104}},\ \bibinfo {pages}
  {103507} (\bibinfo {year} {2021})},\ \Eprint
  {http://arxiv.org/abs/2106.01859} {arXiv:2106.01859 [astro-ph.CO]}
  \BibitemShut {NoStop}%
\bibitem [{\citenamefont {Abbott}\ \emph
  {et~al.}(2017{\natexlab{b}})\citenamefont {Abbott} \emph
  {et~al.}}]{Abbott:2017xzu}%
  \BibitemOpen
  \bibfield  {author} {\bibinfo {author} {\bibfnamefont {B.}~\bibnamefont
  {Abbott}} \emph {et~al.} (\bibinfo {collaboration} {LIGO Scientific, Virgo,
  1M2H, Dark Energy Camera GW-E, DES, DLT40, Las Cumbres Observatory, VINROUGE,
  MASTER}),\ }\href {\doibase 10.1038/nature24471} {\bibfield  {journal}
  {\bibinfo  {journal} {Nature}\ }\textbf {\bibinfo {volume} {551}},\ \bibinfo
  {pages} {85} (\bibinfo {year} {2017}{\natexlab{b}})},\ \Eprint
  {http://arxiv.org/abs/1710.05835} {arXiv:1710.05835 [astro-ph.CO]}
  \BibitemShut {NoStop}%
\bibitem [{\citenamefont {Diaz}\ and\ \citenamefont
  {Mukherjee}(2021)}]{Diaz:2021pem}%
  \BibitemOpen
  \bibfield  {author} {\bibinfo {author} {\bibfnamefont {C.~C.}\ \bibnamefont
  {Diaz}}\ and\ \bibinfo {author} {\bibfnamefont {S.}~\bibnamefont
  {Mukherjee}},\ }\href@noop {} {\  (\bibinfo {year} {2021})},\ \Eprint
  {http://arxiv.org/abs/2107.12787} {arXiv:2107.12787 [astro-ph.CO]}
  \BibitemShut {NoStop}%
\bibitem [{\citenamefont {Soares-Santos}\ \emph {et~al.}(2019)\citenamefont
  {Soares-Santos} \emph {et~al.}}]{Soares-Santos:2019irc}%
  \BibitemOpen
  \bibfield  {author} {\bibinfo {author} {\bibfnamefont {M.}~\bibnamefont
  {Soares-Santos}} \emph {et~al.} (\bibinfo {collaboration} {DES, LIGO
  Scientific, Virgo}),\ }\href {\doibase 10.3847/2041-8213/ab14f1} {\bibfield
  {journal} {\bibinfo  {journal} {Astrophys. J.}\ }\textbf {\bibinfo {volume}
  {876}},\ \bibinfo {pages} {L7} (\bibinfo {year} {2019})},\ \Eprint
  {http://arxiv.org/abs/1901.01540} {arXiv:1901.01540 [astro-ph.CO]}
  \BibitemShut {NoStop}%
\bibitem [{\citenamefont {Del~Pozzo}(2012)}]{DelPozzo:2011yh}%
  \BibitemOpen
  \bibfield  {author} {\bibinfo {author} {\bibfnamefont {W.}~\bibnamefont
  {Del~Pozzo}},\ }\href {\doibase 10.1103/PhysRevD.86.043011} {\bibfield
  {journal} {\bibinfo  {journal} {Phys. Rev.}\ }\textbf {\bibinfo {volume}
  {D86}},\ \bibinfo {pages} {043011} (\bibinfo {year} {2012})},\ \Eprint
  {http://arxiv.org/abs/1108.1317} {arXiv:1108.1317 [astro-ph.CO]} \BibitemShut
  {NoStop}%
\bibitem [{\citenamefont {Zhu}\ \emph {et~al.}(2021)\citenamefont {Zhu},
  \citenamefont {Hu}, \citenamefont {Wang}, \citenamefont {Zhang},
  \citenamefont {Li}, \citenamefont {Hendry},\ and\ \citenamefont
  {Mei}}]{Zhu:2021aat}%
  \BibitemOpen
  \bibfield  {author} {\bibinfo {author} {\bibfnamefont {L.-G.}\ \bibnamefont
  {Zhu}}, \bibinfo {author} {\bibfnamefont {Y.-M.}\ \bibnamefont {Hu}},
  \bibinfo {author} {\bibfnamefont {H.-T.}\ \bibnamefont {Wang}}, \bibinfo
  {author} {\bibfnamefont {J.-D.}\ \bibnamefont {Zhang}}, \bibinfo {author}
  {\bibfnamefont {X.-D.}\ \bibnamefont {Li}}, \bibinfo {author} {\bibfnamefont
  {M.}~\bibnamefont {Hendry}}, \ and\ \bibinfo {author} {\bibfnamefont
  {J.}~\bibnamefont {Mei}},\ }\href@noop {} {\  (\bibinfo {year} {2021})},\
  \Eprint {http://arxiv.org/abs/2104.11956} {arXiv:2104.11956 [astro-ph.CO]}
  \BibitemShut {NoStop}%
\bibitem [{\citenamefont {Messenger}\ and\ \citenamefont
  {Read}(2012)}]{Messenger:2011gi}%
  \BibitemOpen
  \bibfield  {author} {\bibinfo {author} {\bibfnamefont {C.}~\bibnamefont
  {Messenger}}\ and\ \bibinfo {author} {\bibfnamefont {J.}~\bibnamefont
  {Read}},\ }\href {\doibase 10.1103/PhysRevLett.108.091101} {\bibfield
  {journal} {\bibinfo  {journal} {Phys. Rev. Lett.}\ }\textbf {\bibinfo
  {volume} {108}},\ \bibinfo {pages} {091101} (\bibinfo {year} {2012})},\
  \Eprint {http://arxiv.org/abs/1107.5725} {arXiv:1107.5725 [gr-qc]}
  \BibitemShut {NoStop}%
\bibitem [{\citenamefont {Mukherjee}\ \emph
  {et~al.}(2021{\natexlab{a}})\citenamefont {Mukherjee}, \citenamefont
  {Wandelt}, \citenamefont {Nissanke},\ and\ \citenamefont
  {Silvestri}}]{Mukherjee:2020hyn}%
  \BibitemOpen
  \bibfield  {author} {\bibinfo {author} {\bibfnamefont {S.}~\bibnamefont
  {Mukherjee}}, \bibinfo {author} {\bibfnamefont {B.~D.}\ \bibnamefont
  {Wandelt}}, \bibinfo {author} {\bibfnamefont {S.~M.}\ \bibnamefont
  {Nissanke}}, \ and\ \bibinfo {author} {\bibfnamefont {A.}~\bibnamefont
  {Silvestri}},\ }\href {\doibase 10.1103/PhysRevD.103.043520} {\bibfield
  {journal} {\bibinfo  {journal} {Phys. Rev. D}\ }\textbf {\bibinfo {volume}
  {103}},\ \bibinfo {pages} {043520} (\bibinfo {year} {2021}{\natexlab{a}})},\
  \Eprint {http://arxiv.org/abs/2007.02943} {arXiv:2007.02943 [astro-ph.CO]}
  \BibitemShut {NoStop}%
\bibitem [{\citenamefont {Mukherjee}\ \emph
  {et~al.}(2021{\natexlab{b}})\citenamefont {Mukherjee}, \citenamefont
  {Wandelt},\ and\ \citenamefont {Silk}}]{Mukherjee:2020mha}%
  \BibitemOpen
  \bibfield  {author} {\bibinfo {author} {\bibfnamefont {S.}~\bibnamefont
  {Mukherjee}}, \bibinfo {author} {\bibfnamefont {B.~D.}\ \bibnamefont
  {Wandelt}}, \ and\ \bibinfo {author} {\bibfnamefont {J.}~\bibnamefont
  {Silk}},\ }\href {\doibase 10.1093/mnras/stab001} {\bibfield  {journal}
  {\bibinfo  {journal} {Mon. Not. Roy. Astron. Soc.}\ }\textbf {\bibinfo
  {volume} {502}},\ \bibinfo {pages} {1136} (\bibinfo {year}
  {2021}{\natexlab{b}})},\ \Eprint {http://arxiv.org/abs/2012.15316}
  {arXiv:2012.15316 [astro-ph.CO]} \BibitemShut {NoStop}%
\bibitem [{\citenamefont {Heger}\ \emph {et~al.}(2003)\citenamefont {Heger},
  \citenamefont {Fryer}, \citenamefont {Woosley}, \citenamefont {Langer},\ and\
  \citenamefont {Hartmann}}]{Heger:2002by}%
  \BibitemOpen
  \bibfield  {author} {\bibinfo {author} {\bibfnamefont {A.}~\bibnamefont
  {Heger}}, \bibinfo {author} {\bibfnamefont {C.}~\bibnamefont {Fryer}},
  \bibinfo {author} {\bibfnamefont {S.}~\bibnamefont {Woosley}}, \bibinfo
  {author} {\bibfnamefont {N.}~\bibnamefont {Langer}}, \ and\ \bibinfo {author}
  {\bibfnamefont {D.}~\bibnamefont {Hartmann}},\ }\href {\doibase
  10.1086/375341} {\bibfield  {journal} {\bibinfo  {journal} {Astrophys. J.}\
  }\textbf {\bibinfo {volume} {591}},\ \bibinfo {pages} {288} (\bibinfo {year}
  {2003})},\ \Eprint {http://arxiv.org/abs/astro-ph/0212469}
  {arXiv:astro-ph/0212469} \BibitemShut {NoStop}%
\bibitem [{\citenamefont {Farr}\ \emph {et~al.}(2019)\citenamefont {Farr},
  \citenamefont {Fishbach}, \citenamefont {Ye},\ and\ \citenamefont
  {Holz}}]{Farr:2019twy}%
  \BibitemOpen
  \bibfield  {author} {\bibinfo {author} {\bibfnamefont {W.~M.}\ \bibnamefont
  {Farr}}, \bibinfo {author} {\bibfnamefont {M.}~\bibnamefont {Fishbach}},
  \bibinfo {author} {\bibfnamefont {J.}~\bibnamefont {Ye}}, \ and\ \bibinfo
  {author} {\bibfnamefont {D.}~\bibnamefont {Holz}},\ }\href {\doibase
  10.3847/2041-8213/ab4284} {\bibfield  {journal} {\bibinfo  {journal}
  {Astrophys. J. Lett.}\ }\textbf {\bibinfo {volume} {883}},\ \bibinfo {pages}
  {L42} (\bibinfo {year} {2019})},\ \Eprint {http://arxiv.org/abs/1908.09084}
  {arXiv:1908.09084 [astro-ph.CO]} \BibitemShut {NoStop}%
\bibitem [{\citenamefont {Ezquiaga}\ and\ \citenamefont
  {Holz}(2021)}]{Ezquiaga:2020tns}%
  \BibitemOpen
  \bibfield  {author} {\bibinfo {author} {\bibfnamefont {J.~M.}\ \bibnamefont
  {Ezquiaga}}\ and\ \bibinfo {author} {\bibfnamefont {D.~E.}\ \bibnamefont
  {Holz}},\ }\href {\doibase 10.3847/2041-8213/abe638} {\bibfield  {journal}
  {\bibinfo  {journal} {Astrophys. J. Lett.}\ }\textbf {\bibinfo {volume}
  {909}},\ \bibinfo {pages} {L23} (\bibinfo {year} {2021})},\ \Eprint
  {http://arxiv.org/abs/2006.02211} {arXiv:2006.02211 [astro-ph.HE]}
  \BibitemShut {NoStop}%
\bibitem [{\citenamefont {You}\ \emph {et~al.}(2021)\citenamefont {You},
  \citenamefont {Zhu}, \citenamefont {Ashton}, \citenamefont {Thrane},\ and\
  \citenamefont {Zhu}}]{You:2020wju}%
  \BibitemOpen
  \bibfield  {author} {\bibinfo {author} {\bibfnamefont {Z.-Q.}\ \bibnamefont
  {You}}, \bibinfo {author} {\bibfnamefont {X.-J.}\ \bibnamefont {Zhu}},
  \bibinfo {author} {\bibfnamefont {G.}~\bibnamefont {Ashton}}, \bibinfo
  {author} {\bibfnamefont {E.}~\bibnamefont {Thrane}}, \ and\ \bibinfo {author}
  {\bibfnamefont {Z.-H.}\ \bibnamefont {Zhu}},\ }\href {\doibase
  10.3847/1538-4357/abd4d4} {\bibfield  {journal} {\bibinfo  {journal}
  {Astrophys. J.}\ }\textbf {\bibinfo {volume} {908}},\ \bibinfo {pages} {215}
  (\bibinfo {year} {2021})},\ \Eprint {http://arxiv.org/abs/2004.00036}
  {arXiv:2004.00036 [astro-ph.CO]} \BibitemShut {NoStop}%
\bibitem [{\citenamefont {Mastrogiovanni}\ \emph {et~al.}(2021)\citenamefont
  {Mastrogiovanni}, \citenamefont {Leyde}, \citenamefont {Karathanasis},
  \citenamefont {Chassande-Mottin}, \citenamefont {Steer}, \citenamefont
  {Gair}, \citenamefont {Ghosh}, \citenamefont {Gray}, \citenamefont
  {Mukherjee},\ and\ \citenamefont {Rinaldi}}]{Mastrogiovanni:2021wsd}%
  \BibitemOpen
  \bibfield  {author} {\bibinfo {author} {\bibfnamefont {S.}~\bibnamefont
  {Mastrogiovanni}}, \bibinfo {author} {\bibfnamefont {K.}~\bibnamefont
  {Leyde}}, \bibinfo {author} {\bibfnamefont {C.}~\bibnamefont {Karathanasis}},
  \bibinfo {author} {\bibfnamefont {E.}~\bibnamefont {Chassande-Mottin}},
  \bibinfo {author} {\bibfnamefont {D.~A.}\ \bibnamefont {Steer}}, \bibinfo
  {author} {\bibfnamefont {J.}~\bibnamefont {Gair}}, \bibinfo {author}
  {\bibfnamefont {A.}~\bibnamefont {Ghosh}}, \bibinfo {author} {\bibfnamefont
  {R.}~\bibnamefont {Gray}}, \bibinfo {author} {\bibfnamefont {S.}~\bibnamefont
  {Mukherjee}}, \ and\ \bibinfo {author} {\bibfnamefont {S.}~\bibnamefont
  {Rinaldi}},\ }\href {\doibase 10.1103/PhysRevD.104.062009} {\bibfield
  {journal} {\bibinfo  {journal} {Phys. Rev. D}\ }\textbf {\bibinfo {volume}
  {104}},\ \bibinfo {pages} {062009} (\bibinfo {year} {2021})},\ \Eprint
  {http://arxiv.org/abs/2103.14663} {arXiv:2103.14663 [gr-qc]} \BibitemShut
  {NoStop}%
\bibitem [{\citenamefont {Ding}\ \emph {et~al.}(2019)\citenamefont {Ding},
  \citenamefont {Biesiada}, \citenamefont {Zheng}, \citenamefont {Liao},
  \citenamefont {Li},\ and\ \citenamefont {Zhu}}]{Ding:2018zrk}%
  \BibitemOpen
  \bibfield  {author} {\bibinfo {author} {\bibfnamefont {X.}~\bibnamefont
  {Ding}}, \bibinfo {author} {\bibfnamefont {M.}~\bibnamefont {Biesiada}},
  \bibinfo {author} {\bibfnamefont {X.}~\bibnamefont {Zheng}}, \bibinfo
  {author} {\bibfnamefont {K.}~\bibnamefont {Liao}}, \bibinfo {author}
  {\bibfnamefont {Z.}~\bibnamefont {Li}}, \ and\ \bibinfo {author}
  {\bibfnamefont {Z.-H.}\ \bibnamefont {Zhu}},\ }\href {\doibase
  10.1088/1475-7516/2019/04/033} {\bibfield  {journal} {\bibinfo  {journal}
  {JCAP}\ }\textbf {\bibinfo {volume} {04}},\ \bibinfo {pages} {033} (\bibinfo
  {year} {2019})},\ \Eprint {http://arxiv.org/abs/1801.05073} {arXiv:1801.05073
  [astro-ph.CO]} \BibitemShut {NoStop}%
\bibitem [{\citenamefont {Ye}\ and\ \citenamefont
  {Fishbach}(2021)}]{Ye:2021klk}%
  \BibitemOpen
  \bibfield  {author} {\bibinfo {author} {\bibfnamefont {C.}~\bibnamefont
  {Ye}}\ and\ \bibinfo {author} {\bibfnamefont {M.}~\bibnamefont {Fishbach}},\
  }\href {\doibase 10.1103/PhysRevD.104.043507} {\bibfield  {journal} {\bibinfo
   {journal} {Phys. Rev. D}\ }\textbf {\bibinfo {volume} {104}},\ \bibinfo
  {pages} {043507} (\bibinfo {year} {2021})},\ \Eprint
  {http://arxiv.org/abs/2103.14038} {arXiv:2103.14038 [astro-ph.CO]}
  \BibitemShut {NoStop}%
\bibitem [{\citenamefont {Scolnic}\ \emph {et~al.}(2018)\citenamefont {Scolnic}
  \emph {et~al.}}]{Scolnic:2017caz}%
  \BibitemOpen
  \bibfield  {author} {\bibinfo {author} {\bibfnamefont {D.~M.}\ \bibnamefont
  {Scolnic}} \emph {et~al.},\ }\href {\doibase 10.3847/1538-4357/aab9bb}
  {\bibfield  {journal} {\bibinfo  {journal} {Astrophys. J.}\ }\textbf
  {\bibinfo {volume} {859}},\ \bibinfo {pages} {101} (\bibinfo {year}
  {2018})},\ \Eprint {http://arxiv.org/abs/1710.00845} {arXiv:1710.00845
  [astro-ph.CO]} \BibitemShut {NoStop}%
\bibitem [{\citenamefont {Fishbach}\ \emph {et~al.}(2019)\citenamefont
  {Fishbach} \emph {et~al.}}]{Fishbach:2018gjp}%
  \BibitemOpen
  \bibfield  {author} {\bibinfo {author} {\bibfnamefont {M.}~\bibnamefont
  {Fishbach}} \emph {et~al.} (\bibinfo {collaboration} {LIGO Scientific,
  Virgo}),\ }\href {\doibase 10.3847/2041-8213/aaf96e} {\bibfield  {journal}
  {\bibinfo  {journal} {Astrophys. J. Lett.}\ }\textbf {\bibinfo {volume}
  {871}},\ \bibinfo {pages} {L13} (\bibinfo {year} {2019})},\ \Eprint
  {http://arxiv.org/abs/1807.05667} {arXiv:1807.05667 [astro-ph.CO]}
  \BibitemShut {NoStop}%
\bibitem [{\citenamefont {Abbott}\ \emph
  {et~al.}(2021{\natexlab{a}})\citenamefont {Abbott} \emph
  {et~al.}}]{Abbott:2019yzh}%
  \BibitemOpen
  \bibfield  {author} {\bibinfo {author} {\bibfnamefont {B.~P.}\ \bibnamefont
  {Abbott}} \emph {et~al.} (\bibinfo {collaboration} {LIGO Scientific,
  Virgo}),\ }\href {\doibase 10.3847/1538-4357/abdcb7} {\bibfield  {journal}
  {\bibinfo  {journal} {Astrophys. J.}\ }\textbf {\bibinfo {volume} {909}},\
  \bibinfo {pages} {218} (\bibinfo {year} {2021}{\natexlab{a}})},\ \Eprint
  {http://arxiv.org/abs/1908.06060} {arXiv:1908.06060 [astro-ph.CO]}
  \BibitemShut {NoStop}%
\bibitem [{\citenamefont {Vitale}\ \emph {et~al.}(2019)\citenamefont {Vitale},
  \citenamefont {Farr}, \citenamefont {Ng},\ and\ \citenamefont
  {Rodriguez}}]{Vitale:2018yhm}%
  \BibitemOpen
  \bibfield  {author} {\bibinfo {author} {\bibfnamefont {S.}~\bibnamefont
  {Vitale}}, \bibinfo {author} {\bibfnamefont {W.~M.}\ \bibnamefont {Farr}},
  \bibinfo {author} {\bibfnamefont {K.}~\bibnamefont {Ng}}, \ and\ \bibinfo
  {author} {\bibfnamefont {C.~L.}\ \bibnamefont {Rodriguez}},\ }\href {\doibase
  10.3847/2041-8213/ab50c0} {\bibfield  {journal} {\bibinfo  {journal}
  {Astrophys. J. Lett.}\ }\textbf {\bibinfo {volume} {886}},\ \bibinfo {pages}
  {L1} (\bibinfo {year} {2019})},\ \Eprint {http://arxiv.org/abs/1808.00901}
  {arXiv:1808.00901 [astro-ph.HE]} \BibitemShut {NoStop}%
\bibitem [{\citenamefont {Soares De~Souza}\ and\ \citenamefont
  {Sturani}(2021)}]{Mendonca:2019yfo}%
  \BibitemOpen
  \bibfield  {author} {\bibinfo {author} {\bibfnamefont {J.~M.}\ \bibnamefont
  {Soares De~Souza}}\ and\ \bibinfo {author} {\bibfnamefont {R.}~\bibnamefont
  {Sturani}},\ }\href {\doibase 10.1016/j.dark.2021.100830} {\bibfield
  {journal} {\bibinfo  {journal} {Phys. Dark Univ.}\ }\textbf {\bibinfo
  {volume} {32}},\ \bibinfo {pages} {100830} (\bibinfo {year} {2021})},\
  \Eprint {http://arxiv.org/abs/1905.03848} {arXiv:1905.03848 [gr-qc]}
  \BibitemShut {NoStop}%
\bibitem [{\citenamefont {Madau}\ and\ \citenamefont
  {Dickinson}(2014)}]{Madau:2014bja}%
  \BibitemOpen
  \bibfield  {author} {\bibinfo {author} {\bibfnamefont {P.}~\bibnamefont
  {Madau}}\ and\ \bibinfo {author} {\bibfnamefont {M.}~\bibnamefont
  {Dickinson}},\ }\href {\doibase 10.1146/annurev-astro-081811-125615}
  {\bibfield  {journal} {\bibinfo  {journal} {Ann. Rev. Astron. Astrophys.}\
  }\textbf {\bibinfo {volume} {52}},\ \bibinfo {pages} {415} (\bibinfo {year}
  {2014})},\ \Eprint {http://arxiv.org/abs/1403.0007} {arXiv:1403.0007
  [astro-ph.CO]} \BibitemShut {NoStop}%
\bibitem [{\citenamefont {Abbott}\ \emph
  {et~al.}(2021{\natexlab{b}})\citenamefont {Abbott} \emph
  {et~al.}}]{LIGOScientific:2021aug}%
  \BibitemOpen
  \bibfield  {author} {\bibinfo {author} {\bibfnamefont {R.}~\bibnamefont
  {Abbott}} \emph {et~al.} (\bibinfo {collaboration} {LIGO Scientific, VIRGO,
  KAGRA}),\ }\href@noop {} {\  (\bibinfo {year} {2021}{\natexlab{b}})},\
  \Eprint {http://arxiv.org/abs/2111.03604} {arXiv:2111.03604 [astro-ph.CO]}
  \BibitemShut {NoStop}%
\bibitem [{\citenamefont {{LIGO Scientific Collaboration}}(2018)}]{lalsuite}%
  \BibitemOpen
  \bibfield  {author} {\bibinfo {author} {\bibnamefont {{LIGO Scientific
  Collaboration}}},\ }\href {\doibase 10.7935/GT1W-FZ16} {\enquote {\bibinfo
  {title} {{LIGO} {A}lgorithm {L}ibrary - {LALS}uite},}\ }\bibinfo
  {howpublished} {free software (GPL)} (\bibinfo {year} {2018})\BibitemShut
  {NoStop}%
\bibitem [{\citenamefont {Hall}\ and\ \citenamefont
  {Evans}(2019)}]{Hall:2019xmm}%
  \BibitemOpen
  \bibfield  {author} {\bibinfo {author} {\bibfnamefont {E.~D.}\ \bibnamefont
  {Hall}}\ and\ \bibinfo {author} {\bibfnamefont {M.}~\bibnamefont {Evans}},\
  }\href {\doibase 10.1088/1361-6382/ab41d6} {\bibfield  {journal} {\bibinfo
  {journal} {Class. Quant. Grav.}\ }\textbf {\bibinfo {volume} {36}},\ \bibinfo
  {pages} {225002} (\bibinfo {year} {2019})},\ \Eprint
  {http://arxiv.org/abs/1902.09485} {arXiv:1902.09485 [astro-ph.IM]}
  \BibitemShut {NoStop}%
\bibitem [{\citenamefont {Abbott}\ \emph {et~al.}(2019)\citenamefont {Abbott}
  \emph {et~al.}}]{LIGOScientific:2018jsj}%
  \BibitemOpen
  \bibfield  {author} {\bibinfo {author} {\bibfnamefont {B.~P.}\ \bibnamefont
  {Abbott}} \emph {et~al.} (\bibinfo {collaboration} {LIGO Scientific,
  Virgo}),\ }\href {\doibase 10.3847/2041-8213/ab3800} {\bibfield  {journal}
  {\bibinfo  {journal} {Astrophys. J. Lett.}\ }\textbf {\bibinfo {volume}
  {882}},\ \bibinfo {pages} {L24} (\bibinfo {year} {2019})},\ \Eprint
  {http://arxiv.org/abs/1811.12940} {arXiv:1811.12940 [astro-ph.HE]}
  \BibitemShut {NoStop}%
\bibitem [{\citenamefont {Abbott}\ \emph
  {et~al.}(2021{\natexlab{c}})\citenamefont {Abbott} \emph
  {et~al.}}]{LIGOScientific:2020kqk}%
  \BibitemOpen
  \bibfield  {author} {\bibinfo {author} {\bibfnamefont {R.}~\bibnamefont
  {Abbott}} \emph {et~al.} (\bibinfo {collaboration} {LIGO Scientific,
  Virgo}),\ }\href {\doibase 10.3847/2041-8213/abe949} {\bibfield  {journal}
  {\bibinfo  {journal} {Astrophys. J. Lett.}\ }\textbf {\bibinfo {volume}
  {913}},\ \bibinfo {pages} {L7} (\bibinfo {year} {2021}{\natexlab{c}})},\
  \Eprint {http://arxiv.org/abs/2010.14533} {arXiv:2010.14533 [astro-ph.HE]}
  \BibitemShut {NoStop}%
\bibitem [{\citenamefont {Husa}\ \emph {et~al.}(2016)\citenamefont {Husa},
  \citenamefont {Khan}, \citenamefont {Hannam}, \citenamefont {P\"urrer},
  \citenamefont {Ohme}, \citenamefont {Jim\'enez~Forteza},\ and\ \citenamefont
  {Boh\'e}}]{Husa:2015iqa}%
  \BibitemOpen
  \bibfield  {author} {\bibinfo {author} {\bibfnamefont {S.}~\bibnamefont
  {Husa}}, \bibinfo {author} {\bibfnamefont {S.}~\bibnamefont {Khan}}, \bibinfo
  {author} {\bibfnamefont {M.}~\bibnamefont {Hannam}}, \bibinfo {author}
  {\bibfnamefont {M.}~\bibnamefont {P\"urrer}}, \bibinfo {author}
  {\bibfnamefont {F.}~\bibnamefont {Ohme}}, \bibinfo {author} {\bibfnamefont
  {X.}~\bibnamefont {Jim\'enez~Forteza}}, \ and\ \bibinfo {author}
  {\bibfnamefont {A.}~\bibnamefont {Boh\'e}},\ }\href {\doibase
  10.1103/PhysRevD.93.044006} {\bibfield  {journal} {\bibinfo  {journal} {Phys.
  Rev. D}\ }\textbf {\bibinfo {volume} {93}},\ \bibinfo {pages} {044006}
  (\bibinfo {year} {2016})},\ \Eprint {http://arxiv.org/abs/1508.07250}
  {arXiv:1508.07250 [gr-qc]} \BibitemShut {NoStop}%
\bibitem [{\citenamefont {Khan}\ \emph {et~al.}(2016)\citenamefont {Khan},
  \citenamefont {Husa}, \citenamefont {Hannam}, \citenamefont {Ohme},
  \citenamefont {P\"urrer}, \citenamefont {Jim\'enez~Forteza},\ and\
  \citenamefont {Boh\'e}}]{Khan:2015jqa}%
  \BibitemOpen
  \bibfield  {author} {\bibinfo {author} {\bibfnamefont {S.}~\bibnamefont
  {Khan}}, \bibinfo {author} {\bibfnamefont {S.}~\bibnamefont {Husa}}, \bibinfo
  {author} {\bibfnamefont {M.}~\bibnamefont {Hannam}}, \bibinfo {author}
  {\bibfnamefont {F.}~\bibnamefont {Ohme}}, \bibinfo {author} {\bibfnamefont
  {M.}~\bibnamefont {P\"urrer}}, \bibinfo {author} {\bibfnamefont
  {X.}~\bibnamefont {Jim\'enez~Forteza}}, \ and\ \bibinfo {author}
  {\bibfnamefont {A.}~\bibnamefont {Boh\'e}},\ }\href {\doibase
  10.1103/PhysRevD.93.044007} {\bibfield  {journal} {\bibinfo  {journal} {Phys.
  Rev. D}\ }\textbf {\bibinfo {volume} {93}},\ \bibinfo {pages} {044007}
  (\bibinfo {year} {2016})},\ \Eprint {http://arxiv.org/abs/1508.07253}
  {arXiv:1508.07253 [gr-qc]} \BibitemShut {NoStop}%
\bibitem [{\citenamefont {Foreman-Mackey}\ \emph {et~al.}(2013)\citenamefont
  {Foreman-Mackey}, \citenamefont {Hogg}, \citenamefont {Lang},\ and\
  \citenamefont {Goodman}}]{ForemanMackey:2012ig}%
  \BibitemOpen
  \bibfield  {author} {\bibinfo {author} {\bibfnamefont {D.}~\bibnamefont
  {Foreman-Mackey}}, \bibinfo {author} {\bibfnamefont {D.~W.}\ \bibnamefont
  {Hogg}}, \bibinfo {author} {\bibfnamefont {D.}~\bibnamefont {Lang}}, \ and\
  \bibinfo {author} {\bibfnamefont {J.}~\bibnamefont {Goodman}},\ }\href
  {\doibase 10.1086/670067} {\bibfield  {journal} {\bibinfo  {journal} {Publ.
  Astron. Soc. Pac.}\ }\textbf {\bibinfo {volume} {125}},\ \bibinfo {pages}
  {306} (\bibinfo {year} {2013})},\ \Eprint {http://arxiv.org/abs/1202.3665}
  {arXiv:1202.3665 [astro-ph.IM]} \BibitemShut {NoStop}%
\bibitem [{\citenamefont {Ashton}\ \emph {et~al.}(2019)\citenamefont {Ashton}
  \emph {et~al.}}]{Ashton:2018jfp}%
  \BibitemOpen
  \bibfield  {author} {\bibinfo {author} {\bibfnamefont {G.}~\bibnamefont
  {Ashton}} \emph {et~al.},\ }\href {\doibase 10.3847/1538-4365/ab06fc}
  {\bibfield  {journal} {\bibinfo  {journal} {Astrophys. J. Suppl.}\ }\textbf
  {\bibinfo {volume} {241}},\ \bibinfo {pages} {27} (\bibinfo {year} {2019})},\
  \Eprint {http://arxiv.org/abs/1811.02042} {arXiv:1811.02042 [astro-ph.IM]}
  \BibitemShut {NoStop}%
\bibitem [{\citenamefont {Lewis}(2019)}]{Lewis:2019xzd}%
  \BibitemOpen
  \bibfield  {author} {\bibinfo {author} {\bibfnamefont {A.}~\bibnamefont
  {Lewis}},\ }\href {https://getdist.readthedocs.io} {\  (\bibinfo {year}
  {2019})},\ \Eprint {http://arxiv.org/abs/1910.13970} {arXiv:1910.13970
  [astro-ph.IM]} \BibitemShut {NoStop}%
\bibitem [{\citenamefont {Blanchard}\ \emph {et~al.}(2020)\citenamefont
  {Blanchard} \emph {et~al.}}]{Blanchard:2019oqi}%
  \BibitemOpen
  \bibfield  {author} {\bibinfo {author} {\bibfnamefont {A.}~\bibnamefont
  {Blanchard}} \emph {et~al.} (\bibinfo {collaboration} {Euclid}),\ }\href
  {\doibase 10.1051/0004-6361/202038071} {\bibfield  {journal} {\bibinfo
  {journal} {Astron. Astrophys.}\ }\textbf {\bibinfo {volume} {642}},\ \bibinfo
  {pages} {A191} (\bibinfo {year} {2020})},\ \Eprint
  {http://arxiv.org/abs/1910.09273} {arXiv:1910.09273 [astro-ph.CO]}
  \BibitemShut {NoStop}%
\bibitem [{\citenamefont {{Robertson}}\ and\ \citenamefont
  {{Ellis}}(2012)}]{2012ApJ...744...95R}%
  \BibitemOpen
  \bibfield  {author} {\bibinfo {author} {\bibfnamefont {B.~E.}\ \bibnamefont
  {{Robertson}}}\ and\ \bibinfo {author} {\bibfnamefont {R.~S.}\ \bibnamefont
  {{Ellis}}},\ }\href {\doibase 10.1088/0004-637X/744/2/95} {\bibfield
  {journal} {\bibinfo  {journal} {\apj}\ }\textbf {\bibinfo {volume} {744}},\
  \bibinfo {eid} {95} (\bibinfo {year} {2012})},\ \Eprint
  {http://arxiv.org/abs/1109.0990} {arXiv:1109.0990 [astro-ph.CO]} \BibitemShut
  {NoStop}%
\end{thebibliography}
\end{document}